\documentclass[11pt,a4paper,english,oneside]{article}

\usepackage{authblk} % <-- Para gestionar autores y afiliaciones

\usepackage[utf8]{inputenc}
\usepackage[left=2.5cm,top=2.5cm,right=2.5cm,bottom=2cm]{geometry}

\usepackage{amsmath, amssymb, amsthm}

\usepackage{makeidx}
\makeindex

\usepackage{fancyhdr}
\usepackage{graphicx}
\usepackage{subcaption}
\usepackage{epsfig}

\usepackage{multirow}
\usepackage{cancel}
\usepackage{adjustbox}
\usepackage{float}
\usepackage{caption}
\usepackage{hyperref}
\usepackage{quiver}
\usepackage{enumitem}
\usepackage[most]{tcolorbox}
\usepackage{xcolor}
\usepackage{setspace}
\usepackage{parskip}
\usepackage{afterpage}
\usepackage{csquotes}
\usepackage[backend=biber, sorting=none, style=numeric, maxbibnames=99]{biblatex}
\def \punt {\leaders \hbox to 4mm {\hfil . \hfil } \hfill}
\newcommand{\brackets}[1]{\ensuremath{[#1]}}

\newtheorem{thm}{Theorem}[section]

\theoremstyle{definition}
\newtheorem{df}{Definition}[section]

\makeatletter
\newcommand{\skipitems}[1]{%
  \addtocounter{\@enumctr}{#1}%
}
\makeatother

\title{Quantum Machine Learning methods for Fourier-based distribution estimation with application in option pricing}

\author[1,2]{Fernando Alonso}
\author[1,3,*]{Álvaro Leitao}
\author[1,3]{Carlos Vázquez}

\affil[1]{CITIC Research center, Spain}
\affil[2]{Galician Supercomputing Center (CESGA), Spain}
\affil[3]{Department of Mathematics, University of A Coruña, Spain}
\affil[*]{Corresponding author: alvaro.leitao@udc.gal}
\date{\today}

\begin{document}

\maketitle

% REQUIRED
\begin{abstract}
The ongoing progress in quantum technologies has fueled a sustained exploration of their potential applications across various domains. One particularly promising field is quantitative finance, where a central challenge is the pricing of financial derivatives—traditionally addressed through Monte Carlo integration techniques. In this work, we introduce two hybrid classical–quantum methods to address the option pricing problem. These approaches rely on reconstructing Fourier series representations of statistical distributions from the outputs of \textit{Quantum Machine Learning} (QML) models based on \textit{Parametrized Quantum Circuits} (PQCs). We analyze the impact of data size and PQC dimensionality on performance. \textit{Quantum Accelerated Monte Carlo} (QAMC) is employed as a benchmark to quantitatively assess the proposed models in terms of computational cost and accuracy in the extraction of Fourier coefficients. Through the numerical experiments, we show that the proposed methods achieve remarkable accuracy, becoming a competitive quantum alternative for derivatives valuation.
\end{abstract}

% % REQUIRED
% \begin{keywords}
% Quantum Machine Learning, Fourier Series approximation, Parametrized Quantum Circuits, Distribution estimation, Option pricing
% \end{keywords}

% % REQUIRED
% \begin{MSCcodes}
% 65C05, 65R20, 42A10, 81P68
% \end{MSCcodes}

\section{Introduction}

Quantum computing is a field that studies how information processing can be optimized by leveraging the principles of quantum mechanics. Since its inception, significant advances have been made in the design of algorithms and the development of quantum hardware. This has led to a surge in quantum technologies and a continuous search for their applications across various areas. One of these areas is quantitative finance, and the main reason for exploring this field is that some of the internal procedures currently used by financial institutions require highly demanding computations. One of the most relevant applications in finance is the valuation of financial derivatives, which consists of determining the price of a derivative at any date prior to a known maturity date, given by a payoff established in the contract. A derivative is a financial contract, the price of which depends on the future evolution of one or several financial products or rates, which are referred to as underlying. Options represent a main class of derivatives (see \cite{hull}, for a general introduction to options and derivatives).

Several proposals of quantum computing developments for financial applications have been explored in recent years, including the pricing of financial derivatives, as described in some review articles (see \cite{casita}, \cite{bouland2020prospectschallengesquantumfinance} or \cite{Or_s_2019}, for example). Among the possible quantum computing techniques for options pricing, we can find the so-called \textit{Quantum Accelerated Monte Carlo} (QAMC) algorithms (see \cite{Stamatopoulos_2020}, \cite{scriba2025montecarlooptionpricingquantum} or \cite{Udvarnoki_2023}, for example). For these algorithms, the pricing problem is posed in terms of an expectation, which leads to the computation of an integral. In the QAMC approach, this integral is mainly obtained by means of an amplitude estimation algorithm, where the solution of the integral is encapsulated into the amplitude of a quantum state. More precisely, three main steps are involved: a quantum circuit to sample the paths for a prescribed probability distribution, a quantum operator to encode the payoff of the derivative and the aforementioned amplitude estimation routine. This idea has been recently exploited in \cite{manzano_et_al_2025}, where a specific QAMC method proposed in \cite{rqae} is applied to price derivatives with potential negative price. We note that other recent methodologies to calculate integrals with quantum computers are based on the decomposition of the integrand into Fourier series using quantum machine learning models, such as in \cite{Herbert_2022} or \cite{QFIAE}. Alternative formulations for option pricing based on partial differential equations (PDEs) transform the classical PDE for the problem into the propagation governed by an appropriate Hamiltonian operator (see \cite{gonzalez_conde_et_al}, with additional ideas in \cite{fontanela_et_al} and \cite{Kumar_2025}). 

The core ingredient in the present work relies on the area of \textit{Quantum Machine Learning} (QML), from which different proposals have already been made to address financial problems, such as in \cite{leclerc2024}, \cite{thakkar2024} or \cite{wilkens2023}, for example. Among the QML models, one of the most well-known approaches is to construct quantum circuits containing sets of gates, some of which depend on certain parameters that are typically adjusted according to a criterion based on the output of the circuit, thus looking for the set of parameters that provides the optimal value that minimizes a cost function. This classical–quantum concept is known in the literature as \textit{Parametrized Quantum Circuits} (PQCs).  In the context of QML, the ability of PQCs to approximate functions belonging to certain functional spaces has been studied in works like \cite{Schuld_2021} and \cite{P_rez_Salinas_2021}, with positive results when PQCs configurations are expressed in terms generalized trigonometric series.  More recently, in \cite{manzano_et_al_2025} and \cite{manzano2024thesis} a rigorous analysis has been carried out on the approximation of continuous functions and functions in different Sobolev spaces by means these PQCs configurations. In the present article, we recall some of the theorems from \cite{manzano2024thesis} that constitute the theoretical basis of the here proposed QML methodology to approximate the involved probability distributions by means of PQCs. More recently, in \cite{QFIAE} the authors propose a methodology for computing Monte Carlo integrals by decomposing the integrand in trigonometric Fourier series, the coefficients of which are approximated by specific PQCs, namely the so called quantum neuronal networks (QNN). Next, the integration of the trigonometric terms is performed with iterated quantum amplitude estimation techniques (IQAE). Unlike the work in \cite{QFIAE}, in this article we address the Fourier expansion of the payoff of the derivative with the exact coefficients while the probability distribution is obtained from PQCs, that learn the respective Fourier coefficients. Then, the integral estimate can be obtained as the sum of products of coefficients corresponding to the expansion of the two factors of the integral, namely the payoff and the probability distribution. Indeed, for the probability distribution we propose two characterization: the probability density function or the cumulative density function. Moreover, in our financial application, this approach provides the relevant flexibility with great practical interest of getting the information about the probability distribution from samples of the prices of the underlying taken from the market.

Thus, in the present article, a classical–quantum method for the pricing of financial derivatives is proposed and formulated, based on the outputs of QML models built with PQCs, that enables to address the valuation problem pricing from a novel perspective, combining QML tools with fundamental principles of quantitative finance. Within this framework, two different training approaches have been developed to calculate the price of the derivative with the goal of leveraging the expressive power and generalization capabilities that these models offer, in various practical scenarios. Moreover, in order to assess the performance of the designed circuits through numerical results, QAMC is used to obtain a meaningful comparison in terms of both cost and accuracy in the extraction of Fourier coefficients.

The manuscript is organized as follows. In Section \ref{sec2}, we introduce the fundamental concepts that will serve as a foundation throughout the article. More precisely, in Section \ref{subsec21}, an overview of the pricing problem is given; in Section \ref{subsec22}, we demonstrate how, with an appropriate choice of PQC, it is possible to construct Fourier series capable of approximating specific functions; and in Section \ref{subsec23}, a brief overview is provided on how the QAMC tackles the problems written in terms of expectations. Section \ref{sec3} is devoted to the introduction of the newly proposed method based on the use of a QML model built on PQCs. In the first part, Section \ref{subsec31}, the mathematical formulation of the first two methods is presented. In the second part, Section \ref{subsec32}, the implementation of each method is carefully described. Then, in Section \ref{sec4}, the numerical results obtained are presented, specifying the experimental settings in Section \ref{subsec41}, and discussing the impact of the data size and the PQC dimension schemes employed in Section \ref{subsec451} and Section \ref{subsec452}, respectively. Finally, the main conclusions are summarized in Section \ref{sec5}.

\section{Preliminaries}\label{sec2}

\subsection{Basic concepts on options pricing}\label{subsec21}

One of the fundamental areas in quantitative finance concerns the pricing of financial derivatives. A derivative is a financial product whose price depends on the price of another financial product or the value of another financial magnitude (interest rate, exchange rate, etc.) on future dates. This financial magnitude or product is referred to as an underlying factor or underlying asset. In the class of European derivatives, there is usually a unique future date to be considered that is termed the maturity date. Thus, the pricing of financial derivatives consists in determining the price of the derivative at any date prior to its maturity, as the value at maturity is known and given by a payoff established in the contract. Among financial derivatives, which also include forward contracts and futures among others, we will focus on options contracts. European options give the owner the right, but not the obligation, to buy/sell the underlying asset at maturity date or receive the payoff at that date (for example, see \cite{hull} for a review of option types, practical use, and their pricing). Therefore, option pricing requires taking into account the future and uncertain dynamics of the underlying asset price, which is assumed to be a stochastic process (i.e., a random variable at each time). Such dynamics is typically modeled in terms of stochastic differential equations (SDEs). Therefore, the option price is also a stochastic process.

In this section, we introduce the main mathematical concepts and notation to address the option pricing techniques proposed in this article. Let $S_t$ and $v_t$, respectively, denote the prices of the underlying asset and the option at time $t \in [0, T]$, where $T$ is the maturity date. In view of the previous arguments, we can assume that $v_t = V(t,S_t)$, for some function $V:[0,T]\times \mathbb{R} \rightarrow \mathbb{R}$ that relates the option price to time and the underlying asset price. A European option specifies the payoff $h$ received by its holder at maturity date, which depends on the underlying asset price at maturity date:
\begin{equation*}
    v_T = h(T,S_T).
\end{equation*}

Using option pricing theory (see \cite{hull}, for example), the price of a European option at time $t<T$ can be obtained in terms of a conditional expectation in the form:
\begin{equation}
v_t = e^{-r(T - t)} \mathbb{E}^{\mathbb{Q}} \left[ h(T,S_T) \mid \mathcal{F}_t \right] = e^{-r(T - t)} \int_{\mathbb{R}} h(T, y) ~ f(y \mid \mathcal{F}_t) \, dy,
\label{esper}
\end{equation}
where $\mathbb{E}^{\mathbb{Q}}$ denotes the expectation under a probability measure $\mathbb{Q}$\footnote{In the literature, it is common to refer to it as the risk-neutral measure.}, $r$ is the constant risk-free interest rate and $\mathcal{F}_t$ denotes the $\sigma$-algebra containing the market information available up to time $t$, which is assumed to be known. Moreover, $f(\cdot)$ is the probability density function (PDF) of the asset price process $S_t$ under $\mathbb{Q}$. 

Note that the expression \eqref{esper} indicates that the value of the derivative at time $t$ is the discounted price of the expected payoff, conditional on the market information available up to time $t$. Moreover, in view of expression \eqref{esper}, the computation of the expectation to obtain the option price requires appropriate integration methods for general payoff and PDF expressions. In the present research work we mainly aim to take advantage of QML techniques to estimate the PDF and approximate this integral.

\subsection{PQCs as universal approximators}\label{subsec22}

A very common approach within the quantum-classical framework of QML consists of using trainable quantum circuits as models, in a similar way to neural networks. In this approach, quantum gates are used both to encode the data inputs, $x = (x_1, \ldots, x_N)$, and to implement trainable weights, $\boldsymbol{\theta} = (\theta_1, \ldots, \theta_M)$. The circuit is measured multiple times to estimate the expected value of an observable, and this result is understood as a prediction, leading to the implementation of a function $f_\theta(x)$. This approach is referred by various authors in the literature as PQCs (see for example \cite{liu2024analysisparameterizedquantumcircuits}, \cite{Benedetti_2019} or \cite{Ostaszewski2021structure}). The information extracted from the circuit and the process of evaluating the cost function are both classical, which makes PQC-based algorithms hybrid.

\begin{figure}[h]
\centering
\includegraphics[width=0.6\textwidth]{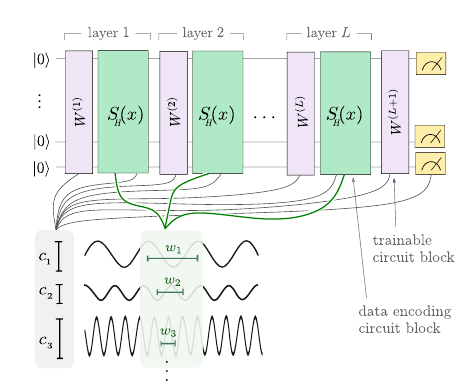}
\caption{Scheme of a quantum circuit composed of $L$ layers, where each layer consists of a trainable circuit block $W^{(i)}(\boldsymbol{\theta})$, with $i \in \{1, \ldots, L\}$, and a data-encoding block $S_H(x)$, from \cite{Schuld_2021}.}
\label{figschultz}
\end{figure}

Let a univariate quantum model be defined as the expectation value of an observable with respect to a state prepared by a PQC, that is:
\begin{equation*}
f_\theta(x) = \langle 0 | U^\dagger(x,\boldsymbol{\theta}) M U(x,\boldsymbol{\theta}) | 0 \rangle,
\end{equation*}
where $|0\rangle$ is one of the computational basis states, $U(x,\boldsymbol{\theta})$ is a quantum circuit that depends on the input ($x$) and on a set of parameters ($\boldsymbol{\theta}$), and $M$ is an observable. The quantum circuit representing our model will be constructed from $L$ layers, each one consisting of a data-encoding block $ S_H(x)$ and a trainable block $W(\boldsymbol{\theta})$ controlled by the parameters $\boldsymbol{\theta}$, as shown in Figure \ref{figschultz}. For simplicity, in the subsequent developments, we will assume that the trainable blocks are arbitrary unitary operations, i.e., $W(\boldsymbol{\theta}) = W$, and we will omit the subscript in $f_{\boldsymbol{\theta}}$ hereafter. Thus, the total quantum circuit has the form
\begin{equation}
U(x) = W^{(L+1)}(\boldsymbol{\theta}) S_H(x) W^{(L)}(\boldsymbol{\theta}) \cdots W^{(2)}(\boldsymbol{\theta}) S_H(x) W^{(1)}(\boldsymbol{\theta}),
\end{equation}
where the data-encoding block is identical in each layer and has the form
%\begin{equation*}
    $S_H(x) = e^{-x_1 H} \otimes \cdots \otimes e^{-x_N H}$,
%\end{equation*}
for $H$ a Hamiltonian that generates the time evolution used to encode the data.

In this context, the fundamental idea under the developments we will carry out is that, given an appropriate choice of circuits, the quantum function representing our model can be written as a Fourier series of the form
\begin{equation*}
f_\theta(x) = \sum_{\omega \in \Omega} c_\omega(\theta) e^{i \omega x}.
\end{equation*}
Several references in the literature prove that the frequency spectrum ($\Omega \subseteq \mathbb{R}^N$) is completely determined by the eigenvalues of the Hamiltonians forming the encoding block, while the complete circuit design controls the coefficients ($c_\omega$) that a quantum model can adjust (for example, see  \cite{Schuld_2021} and \cite{P_rez_Salinas_2021}). Moreover, in many cases the frequencies are integers ($\Omega \subseteq \mathbb{Z}^N$), and the sum becomes a partial Fourier series\footnote{The term \textit{partial Fourier series} indicates that only a subset of the Fourier coefficients are nonzero.}
\begin{equation*}
f_\theta(x) = \sum_{n \in \Omega} c_n(\theta) e^{i n x},
\end{equation*}
where the functions $e^{i n x}$ form an orthogonal basis. This formalism allows quantum models to be studied using the tools of Fourier analysis.

From the aforementioned results, it arises the intuition that, employing a sufficiently large number of repetitions of encoding gates or Hamiltonians with sufficiently large dimension and suitably non-degenerate spectrum, quantum models can approximate a wide range of functions.

Thus two key aspects appear: the \emph{universality} and the \emph{expressivity} of PQCs, that is, the ability of PQCs to approximate any function belonging to a given function space defined over a prescribed domain, up to arbitrary precision with respect to a specific norm. This idea was first explored in \cite{Schuld_2021} and \cite{P_rez_Salinas_2021}, and following these developments it is shown that, if trainable circuit blocks are allowed with sufficient flexibility to implement arbitrary global unitary gates, then there exists a quantum model with $L = 1$ capable of realizing any possible set of Fourier coefficients.

Furthermore, in \cite{Manzano_2025} and \textit{Chapter} 5 of \cite{manzano2024thesis} it has been proven that PQCs can approximate arbitrarily well the space of continuous functions, the space of $p$-integrable functions and the Sobolev space $H^k$, which is the set of functions whose derivatives up to order $k$ are $L^2$-integrable. Before recalling these results, it is necessary to introduce the following definition of universal Hamiltonian family proposed in \cite{Schuld_2021}.

\begin{df}\label{def:hamiltonianfamily}
Let $\{H_{m} \mid m \in \mathbb{N}\}$ be a Hamiltonian family where $H_{m}$ acts on $m$ subsystems
of dimension $d$.
Such a Hamiltonian family gives rise to a family of models $\{f_{m}\}$ in the following way:
\begin{equation}
f_{m}(x) = \langle \Gamma \,|\, S^{\dagger}_{H_m}(x) M S_{H_m}(x) \,|\, \Gamma \rangle
\label{quantum_model_family}.
\end{equation}
Furthermore, for each $m \in \mathbb{N}$, the set
%\begin{equation*}
$\Omega_{H_{m}} := 
\bigl\{ \lambda^{m}_{j} - \lambda^{m}_{k} 
\;\big|\;
j,k \in \{1,\dots,d_{m}\} \bigr\}$,
%\end{equation*}
where $\{\lambda^{m}_{1},\dots,\lambda^{m}_{d_{m}}\}$ are the eigenvalues of $H_{m}$, 
is defined as the \emph{frequency spectrum} of $H_{m}$. A Hamiltonian family is a universal Hamiltonian family if for any $Z\in \mathbb{N}$ there exists $m\in \mathbb{N}$ such that $\{-Z,\dots,0,\dots,Z\} \subset H_m(\Omega)$.
\end{df}

Based on the previous definition, in \textit{Chapter} 5 of \cite{manzano2024thesis}, the following three theorems have been proved.

\begin{thm}{\textbf{(Convergence in $C^0$)}}

Let $\{H_m\}$ be a universal Hamiltonian family,
and $\{f_m\}$ the associated quantum model family (\ref{quantum_model_family}).
For all functions $f^{\ast} \in C_{0}(U)$ where $U$ is compactly contained 
in $[0,2\pi]^N$, and for all $\epsilon > 0$, 
there exists some $m' \in \mathbb{N}$, some state 
$\lvert \Gamma \rangle \in \mathbb{C}^{d_{m'}}$, and some observable $M$ such that 
$f_{m'}$ converges uniformly to $f^{\ast}$:
\begin{equation*}
\| f_{m'} - f^{\ast} \|_{C^{0}} < \epsilon.
\end{equation*}
with
\begin{equation*}
\| f_{m'} - f^{\ast} \|_{C^{0}} := 
\sup_{x \in [0,2\pi]^N} 
\| f_{m'}(x) - f^{\ast}(x) \|.
\end{equation*}
\end{thm}

\begin{thm}{\textbf{(Convergence in $L^p$)}}

Let $\{H_m\}$ be a universal Hamiltonian family, 
and $\{f_m\}$ the associated quantum model family (\ref{quantum_model_family}).
For all functions $f^{\ast} \in L^{p}\!\bigl([0,2\pi]^N\bigr)$
where $1 \leq p < \infty$, and for all $\epsilon > 0$, 
there exists some $m' \in \mathbb{N}$, some state 
$\lvert \Gamma \rangle \in \mathbb{C}^{d_{m'}}$, and some observable $M$ such that:
\begin{equation*}
\| f_{m'} - f^{\ast} \|_{L^p} < \epsilon,
\end{equation*}
with
\begin{equation*}
\| f_{m'} - f^{\ast} \|_{L^p} := \left( \int_{[0,2\pi]^N} |f_{m'} - f^{\ast}|^p \, dP \right)^{1/p}.
\end{equation*}
\end{thm}

\begin{thm}{\textbf{(Convergence in $H^{k}$)}}

Let $\{H_m\}$ be a universal Hamiltonian family
and $\{f_m\}$ the associated quantum model family (\ref{quantum_model_family}).
For all functions $f^{\ast} \in H^{k+1}(U)$ where $U$ is compactly contained 
in the closed cube $[0,2\pi]^N$, and for all $\epsilon > 0$, 
there exists some $m' \in \mathbb{N}$, some state 
$\lvert \Gamma \rangle \in \mathbb{C}^{d_{m'}}$, and some observable $M$ such
that $f_{m'}$ converges to $f^{\ast}$ with respect to the $H^{k}$–norm:
\begin{equation*}
\| f_{m'} - f^{\ast} \|_{H^{k}} < \epsilon,
\end{equation*}
with
\begin{align*}
\| f_{m'} - f^{\ast} \|_{H^k} := \left( \sum_{|\alpha| \leq k} \int_U \Bigg|\frac{\partial^{|\alpha|}}
{\partial x_{1}^{\alpha_{1}} \cdots \partial x_{N}^{\alpha_{N}}}(f_{m'} - f^{\ast})(x)\Bigg|^2 \, dP(x) \right)^{1/2}.
\end{align*}
\end{thm}

As discussed in \textit{Chapter} 6 of \cite{manzano2024thesis}, the relevance of these results is that the generalization bounds of the empirical risk defined in $H^{1}$ imply the minimization of the empirical risk in $C^{0}$. This is really strong because the convergence in the sense of $C^{0}$ is equivalent to a convergence of every point to the true solution.

\subsection{Quantum accelerated Monte Carlo techniques}\label{subsec23}

Monte Carlo method is one of the best-known integration techniques for solving option pricing problems, when formulated in terms of expectations. This method gives an approximation of the value of definite integrals by generating random samples within the integration region and computing the average value of the function evaluated in these samples \cite{glasserman}.

Let us consider the definite integral
\begin{equation*}
\mathbb{E}[g(x)] =\int_{x_{\text{min}}}^{x_{\text{max}}} g(x)f(x)\,dx,
\end{equation*}
where $f$ is a PDF with support contained in the interval $[x_{\text{min}}, x_{\text{max}}]$ and $g$ is a function of interest, namely, the payoff in the options pricing problem. The Monte Carlo method consists in generating $I$ independent and identically distributed samples $x_i$, for $i = 0, \dots, I-1$, drawn from the PDF $f$, such that the value of the integral is approximated by 
\begin{equation*}
\int_{x_{\text{min}}}^{x_{\text{max}}} g(x)f(x)\,dx \approx \sum_{i=0}^{I - 1} g(x_i)f(x_i).
\end{equation*}
Since this method can be computationally demanding for certain types of integrals, in recent years the advantages offered by quantum computing have been exploited to develop the QAMC method \cite{Montanaro_2015}, which achieves a quadratic improvement, in terms of the mean squared error, in the number of queries required compared to its classical counterpart. 

The idea behind this method is to encapsulate the value of the expectation within the amplitudes of a quantum state, and then maximize the probability of obtaining 
this value when performing a measurement. For this purpose, the following state is constructed
\begin{equation*}
	|\psi\rangle = \sum_{x=0}^{2^n-1} \sqrt{g(x)f(x)} \, |x\rangle^n |1\rangle 
	+ \sum_{x=0}^{2^n-1} \sqrt{(1 - g(x))f(x)} \, |x\rangle^n |0\rangle.
\end{equation*}

If we now conveniently define the following quantities
\begin{eqnarray*}
	q &=& \sum_{x=0}^{2^n-1} g(x) f(x) \approx \mathbb{E}[g(x)] \\ 
	|\tilde{\psi}_1\rangle &=& \frac{1}{\sqrt{q}} \sum_{x=0}^{2^n-1} \sqrt{g(x)} \sqrt{f(x)} \, |x\rangle^n, \\
	|\tilde{\psi}_0\rangle &=& \frac{1}{\sqrt{1-q}} \sum_{x=0}^{2^n-1} \sqrt{1 - g(x)} \sqrt{f(x)} \, |x\rangle^n, 
\end{eqnarray*}
and rewrite $|\psi\rangle$ as
\begin{equation*}
	|\psi\rangle = \sqrt{q} \, |\tilde{\psi}_1\rangle |1\rangle 
	+ \sqrt{1-q} \, |\tilde{\psi}_0\rangle |0\rangle,
\end{equation*}
we can observe that the problem can be solved through \textit{Quantum Amplitude Estimation} (QAE) \cite{qaeorg}, since the probability 
of the state $a$ corresponds to the integral to be computed, that estimates $\mathbb{E}[g(x)]$.

However, QAMC presents some drawbacks for which several modifications of its original formulation have been proposed. For instance, to address the problem that emerges from the \textit{Quantum Phase Estimation} (QPE) \cite{variousmonte} subroutine, alternatives such as the \textit{Iterative Quantum Amplitude Estimation} (IQAE) \cite{IQAE} or the \textit{Real Quantum Amplitude Estimation} (RQAE) \cite{rqae} have been developed. Furthermore, to adequately prepare the initial quantum state different ideas have been proposed, as discussed in \cite{Carrera_Vazquez_2021}, \cite{holmes2020efficientquantumcircuitsaccurate}, and \cite{grover2002creatingsuperpositionscorrespondefficiently}. Other approaches include methods based on decomposing the integrand into Fourier series, such as \textit{Fourier Quantum Monte Carlo Integration} (FQMCI) \cite{Herbert_2022} or \textit{Quantum Fourier Iterative Amplitude Estimation} (QFIAE) \cite{QFIAE}, which propose ways to compute the Fourier series using QML.

\section{Formulation and methodology}\label{sec3}

As previously mentioned, in quantitative finance, efficient numerical methods are required to value complex contracts and calibrate various financial models. Existing methods can be classified into three main groups: numerical methods for partial differential equations, Monte Carlo simulation techniques, and numerical integration methods, each presenting its own advantages and disadvantages depending on the specific financial application. \

As mentioned before, the rise of quantum computing can lead to a new set of methods which can accelerate numerical simulations and potentially achieve more efficient valuation and calibration of complex financial models. In this work, we explore such potential through classical–quantum techniques built on PQC-based QML models. The following sections describe the formulation and implementation of these techniques, along with the methodology

\subsection{Fourier series approximation of PDF and CDF}\label{subsec31}

The here proposed approaches rely on the use of trigonometric Fourier series, that is, series based on sine and cosine functions. They rely on the same idea, although their main difference is that the first one uses the trigonometric series of the PDF, as in \cite{cosine}, and the second one uses the trigonometric series of the cumulative distribution function (CDF), as in \cite{andersenh}.

\subsubsection{PDF approximation formulation}

Returning to the formulation \eqref{esper} of the derivative pricing problem for European vanilla options, the starting point lies in the calculation of the pricing formula under the risk-neutral measure, given by
\begin{equation}
V(t_0,x) = e^{-r(T-t_0)} \, \mathbb{E}^{\mathbb{Q}}[h(T,y) | x] = e^{-r (T-t_0)} \int_{\mathbb{R}} h(T,y) f(y | x) \, dy,
\label{option_price}
\end{equation}
where $x$ and $y$ are the state variables at times $t_0$ and $T$, respectively, and $f(y | x)$ is the probability density of $y$ conditioned on $x$.

In finance, it is common to work in practice with PDFs whose tails tend to vanish, so it can be assumed that there exists an interval $[a, b] \subset \mathbb{R}$ such that the integral in (\ref{option_price}) can be approximated without significant loss of accuracy, i.e.,
\begin{equation}
V(t_0,x) \approx e^{-r (T-t_0)} \int_a^b h(T,y) f(y | x) \, dy.
\label{formulamlaprecio1}
\end{equation}
Since $f(y | x)$ is usually not known explicitly, we can approximate the density by its $\mathcal{K}$-truncated trigonometric Fourier series expansion in $y$
\begin{equation}
    f(y | x) \approx  \frac{A_0^f}{2} + \sum_{k=1}^{\mathcal{K}} \left( A_k^f \cos\left(2\pi k\frac{(y - a)}{b - a}\right) + B_k^f \sin\left(2\pi k\frac{(y - a)}{b - a}\right) \right),
    \label{serieprob}
\end{equation}
where
% \begin{equation*}
% A_k^f = \frac{2}{b-a} \int_{a}^{b} f(y | x) \cos\left(2\pi k\frac{(y - a)}{b - a}\right) \, dy \quad \text{and} \quad B_k^f = \frac{2}{b-a} \int_{a}^{b} f(y | x) \sin\left(2\pi k\frac{(y - a)}{b - a}\right) \, dy.
% \end{equation*}
\begin{equation*}
A_k^f = \frac{2}{b-a} \int_{a}^{b} f(y | x) \cos\left(2\pi k\frac{(y - a)}{b - a}\right) \, dy,
\end{equation*}
and
\begin{equation*}
B_k^f = \frac{2}{b-a} \int_{a}^{b} f(y | x) \sin\left(2\pi k\frac{(y - a)}{b - a}\right) \, dy.    
\end{equation*}
It should be remarked that regarding the underlying asset price process, we can use the same arguments as in \cite{cosine} to ensure that, due to the conditions required for the existence of the Fourier series, it is possible to truncate the number of terms in the series while controlling the accuracy.

By substituting equation (\ref{serieprob}) into equation (\ref{formulamlaprecio1}), we obtain
% \begin{equation}\label{approxi}
% V(t_0,x) \approx e^{-r (T-t_0)} \int_a^b h(T,y) \left(\frac{A_0^f}{2} + \sum_{k=1}^{\mathcal{K}} \left( A_k^f \cos\left(2\pi k\frac{(y - a)}{b - a}\right) + B_k^f \sin\left(2\pi k\frac{(y - a)}{b - a}\right) \right) \right) \, dy.
% \end{equation}
\begin{equation}\label{approxi}
\begin{aligned}
V(t_0,x) &\approx e^{-r (T-t_0)} \int_a^b h(T,y) \left(\frac{A_0^f}{2} + \sum_{k=1}^{\mathcal{K}}  A_k^f \cos\left(2\pi k\frac{(y - a)}{b - a}\right) \right) \, dy \\
&+ e^{-r (T-t_0)} \int_a^b h(T,y)  \sum_{k=1}^{\mathcal{K}} B_k^f \sin\left(2\pi k\frac{(y - a)}{b - a}\right)  \, dy.
\end{aligned}
\end{equation}
Next, if we now exchange the summation and the integral in (\ref{approxi}) and introduce the definitions
% \begin{equation*}
% C_k := \frac{2}{b - a} \int_a^b h(T,y) \cos\left( 2\pi k\frac{(y - a)}{b - a} \right) \, dy, \quad \text{and} \quad D_k := \frac{2}{b-a} \int_{a}^{b} h(T,y) \sin\left(2\pi k\frac{(y - a)}{b - a}\right) \, dy,
% \end{equation*}
\begin{equation*}
C_k := \frac{2}{b - a} \int_a^b h(T,y) \cos\left( 2\pi k\frac{(y - a)}{b - a} \right) \, dy,
\end{equation*}
and
\begin{equation*}
D_k := \frac{2}{b-a} \int_{a}^{b} h(T,y) \sin\left(2\pi k\frac{(y - a)}{b - a}\right) \, dy,
\end{equation*}
we obtain
\begin{equation*}
V(t_0,x) \approx \frac{1}{2}(b - a) e^{-r\Delta t}\left( \frac{A_0^f \cdot C_0}{2} + \sum_{k=1}^\mathcal{K} \left( A_k^f \cdot C_k + B_k^f \cdot D_k \right) \right).
\end{equation*}

It should be noted that $C_k$ and $D_k$ are the coefficients of the trigonometric Fourier series of $h(T,y)$. Thus, the integral of the product of two real functions, $f(y| x)$ and $h(T,y)$ has been transformed into the sum of the product of the respective coefficients of their trigonometric Fourier series.

\subsubsection{CDF approximation formulation}

In the original formulation of the valuation integral, the integrand may have either infinite or bounded support, and the payoff function may be only piecewise smooth and have discontinuities. Therefore, working directly with the PDF can be numerically unstable or ill-posed. Unlike the PDF, the CDF provides a smoother and continuous representation of the underlying variable's behavior regardless its critical features, which allows to efficiently handle discontinuities and to come up with more stable numerical approximations.

Firstly, assuming that the derivative of the payoff function has a discontinuity at $c \in [a,b]$, we split valuation integral as
% \begin{equation*}
% V(t_0,x) = e^{-r (T-t_0)} \int_a^b h(T,y) f(y| x) \, dy = e^{-r (T-t_0)} \Big( \int_a^c h(T,y) f(y| x) \, dy + \int_c^b h(T,y) f(y| x) \, dy \Big).
% \end{equation*}
\begin{equation*}
\begin{aligned}
V(t_0,x) &= e^{-r (T-t_0)} \int_a^b h(T,y) f(y| x) \, dy \\
&= e^{-r (T-t_0)} \left( \int_a^c h(T,y) f(y| x) \, dy + \int_c^b h(T,y) f(y| x) \, dy \right).
\end{aligned}
\end{equation*}
Then, integrating by parts we obtain
% \begin{equation}\label{formparts}
% V(t_0,x) = e^{-r (T-t_0)} \left( h(T,b)F(b) - h(T,a)F(a) - \int_a^c h'(T,y) F(y) \, dy - \int_c^b h'(T,y) F(y) \, dy \right),
% \end{equation}
\begin{equation}\label{formparts}
\begin{aligned}
V(t_0,x) &= e^{-r (T-t_0)} \left( h(T,b)F(b) - h(T,a)F(a) \right) \\
&- e^{-r (T-t_0)} \left(\int_a^c h'(T,y) F(y) \, dy + \int_c^b h'(T,y) F(y) \, dy \right),
\end{aligned}
\end{equation}
where the CDF is given by 
\begin{equation*}
F(y) = \int_{-\infty}^y f(x) \, dx.
\end{equation*}
Following the same reasoning as before, it is possible to define the Fourier series of period $2(b-a)$ in an interval $[\hat{a}, \hat{b}]$, whose election will be later motivated. Thus, the Fourier series of the CDF is
\begin{equation}
F(y) \approx \frac{A_0^F}{2} + \sum_{k=1}^{\mathcal{K}} \left( A_k^F \cos\left(2\pi k\frac{(y - \hat{a})}{\hat{b} - \hat{a}}\right) + B_k^F \sin\left(2\pi k\frac{(y - \hat{a})}{\hat{b} - \hat{a}}\right) \right).
\label{cdfseries}
\end{equation}
Substituting  (\ref{cdfseries}) in (\ref{formparts}) and exchanging the summation with the integral we obtain
% \begin{align*}
% V(t_0,x) \approx e^{-r\Delta t} \Bigg(  h(T,b)F(b) - h&(T,a)F(a) - \Big( \frac{A_0^F \cdot C_0^a}{2} + \sum_{k=1}^\mathcal{K} \left( A_k^F \cdot C_k^a + B_k^F \cdot D_k^a \right) \Big) - \\ &- \Big( \frac{A_0^F \cdot C_0^b}{2} + \sum_{k=1}^\mathcal{K} \left( A_k^F \cdot C_k^b + B_k^F \cdot D_k^b \right) \Big) \Bigg),
% \end{align*}
\begin{equation*}
\begin{aligned}
V(t_0,x) &\approx e^{-r\Delta t} \left(  h(T,b)F(b) - h(T,a)F(a) \right) \\
&- e^{-r\Delta t} \left( \frac{A_0^F \cdot C_0^a}{2} + \sum_{k=1}^\mathcal{K} \left( A_k^F \cdot C_k^a + B_k^F \cdot D_k^a \right) \right) \\
&- e^{-r\Delta t} \left( \frac{A_0^F \cdot C_0^b}{2} + \sum_{k=1}^\mathcal{K} \left( A_k^F \cdot C_k^b + B_k^F \cdot D_k^b \right) \right)
\end{aligned}
\end{equation*}
where
% \begin{equation}\label{eq:Ck_Dk_a}
% C_k^a := \int_a^c h'(T,y) \cos\left( 2\pi k\frac{(y - \hat{a})}{\hat{b} - \hat{a}} \right) \, dy, \quad \text{and} \quad D_k^a := \int_a^c h'(T,y) \sin\left(2\pi k\frac{(y - \hat{a})}{\hat{b} - \hat{a}}\right) \, dy,
% \end{equation}
% \begin{equation}\label{eq:Ck_Dk_b}
% C_k^b := \int_c^b h'(T,y) \cos\left( 2\pi k\frac{(y - \hat{a})}{\hat{b} - \hat{a}} \right) \, dy, \quad \text{and} \quad D_k^b := \int_c^b h'(T,y) \sin\left(2\pi k\frac{(y - \hat{a})}{\hat{b} - \hat{a}}\right) \, dy.
% \end{equation}
\begin{equation}\label{eq:Ck_Dk}
\begin{aligned}
C_k^a &:= \int_a^c h'(T,y) \cos\left( 2\pi k\frac{(y - \hat{a})}{\hat{b} - \hat{a}} \right)dy, \,\, D_k^a := \int_a^c h'(T,y) \sin\left(2\pi k\frac{(y - \hat{a})}{\hat{b} - \hat{a}}\right) dy, \\
C_k^b &:= \int_c^b h'(T,y) \cos\left( 2\pi k\frac{(y - \hat{a})}{\hat{b} - \hat{a}} \right)dy, \,\, D_k^b := \int_c^b h'(T,y) \sin\left(2\pi k\frac{(y - \hat{a})}{\hat{b} - \hat{a}}\right)dy.
\end{aligned}
\end{equation}
It should be remarked that, in contrast with the previous method, the quantities defined in \eqref{eq:Ck_Dk} do not correspond to the Fourier coefficients of the series that approximates $h(T,y)$, because the integration domain does not match with the one of the basis functions.

\subsection{Methodology}\label{subsec32}

As mentioned before, the proposed technique is based on the approximation of functions through Fourier series extracted from the QML models built on PQCs.
This approximation allows for a flexible capture of the functional structure of the distributions involved in the pricing models, taking advantage of the expressive and differentiable capabilities of these models. Moreover, the integration of classical and quantum computing provides an alternative framework for the valuation of derivatives and the approximation of complex financial functions. For this purpose, three approaches have been designed.

In the first one, supervised learning is employed, using datasets that contain both inputs and labeled outputs. In this case, a PQC is trained to approximate the PDF of the variable representing the underlying asset price and to extract its Fourier series. The coefficients corresponding to the payoff of the derivative are obtained analytically.

The second approach is more realistic from a practical point of view, since in derivatives pricing one rarely has access to the exact PDF of the underlying asset price, but to asset price's evolution in time. Therefore, in this second case, self-supervised learning is employed, after providing the model with a sufficiently representative set of asset samples. From these samples, the model must infer the implicit distribution and estimate the coefficients needed to calculate the price of the derivative.

Note that, as argued in \cite{manzano2024thesis}, in both cases it is assumed that the probability distribution $P(x, y)$ yields a deterministic mapping for some function $g^* : \mathcal{X} \rightarrow \mathcal{Y}$. %\subseteq H^1(\mathcal{X})$.  
Therefore, to solve the classical problem
\begin{equation*}
    g = \arg \min_{\hat{g} \in \mathcal{M}} R(\hat{g}),
\end{equation*}
for $\mathcal{M}$ a subset of functions in some functional space. Instead of working with the joint probability distribution $P(x, y)$, so that
\begin{equation*}
R(g) = \int_{\mathcal{X} \times \mathcal{Y}} \ell(g^*(x), g(x)) \, dP(x,y),
\end{equation*}
it is possible to work with the marginal distribution $P_\mathcal{X} = P_\mathcal{X}(x)$, so that
\begin{equation*}
R(g) = \int_\mathcal{X} \ell(g^*(x), g(x)) \, dP_\mathcal{X}(x).
\end{equation*}
Since in both scenarios we will work in the Sobolev space $H^1(\mathcal{X})$ where $\mathcal{X} \subset \mathbb{R}$ and $g^* \in \mathcal{T}\subseteq H^1(\mathcal{X})$, the corresponding risk can be defined in terms of the usual norm in $H^1$ as
\begin{equation*}
R_{H^1}(g) = \|g^* - g\|_{H^1}^2 = 
\int_\mathcal{X} \left( g^*(x) - g(x) \right)^2  +
\left( \frac{\partial g^*(x)}
{\partial x}  - \frac{\partial g(x)}
{\partial x}  \right)^2 \, dP_X(x),
\end{equation*}
and the empirical risk can be defined as
\begin{align}
R_{h^1}^S(g^*, g) &=
\frac{1}{I} \sum_{i=0}^{I-1} \left( g^*(x_i) - g(x_i) \right)^2 + \frac{1}{I} \sum_{i=0}^{I-1} 
\left( \frac{\partial g^*}{\partial x}(x_i) - \frac{\partial g}{\partial x}(x_i) \right)^2,
\label{ref_risk_generico}
\end{align}
where $I$ is the number of samples considered in the corresponding dataset. The use of information about the derivatives of the outputs with respect to the inputs was first introduced in \cite{hugediff}, giving rise to a new subfield of machine learning known as \textit{Differential Machine Learning} (DML). In accordance with the aforementioned theorems, several results show that incorporating this information can significantly improve a model’s training performance, as it forces the function estimates to converge point-wise rather than on average.

\begin{figure}[H]
    \centering
     \subfloat[Training in $\brackets{-\pi, \pi}$]{\includegraphics[width = 0.465\textwidth]{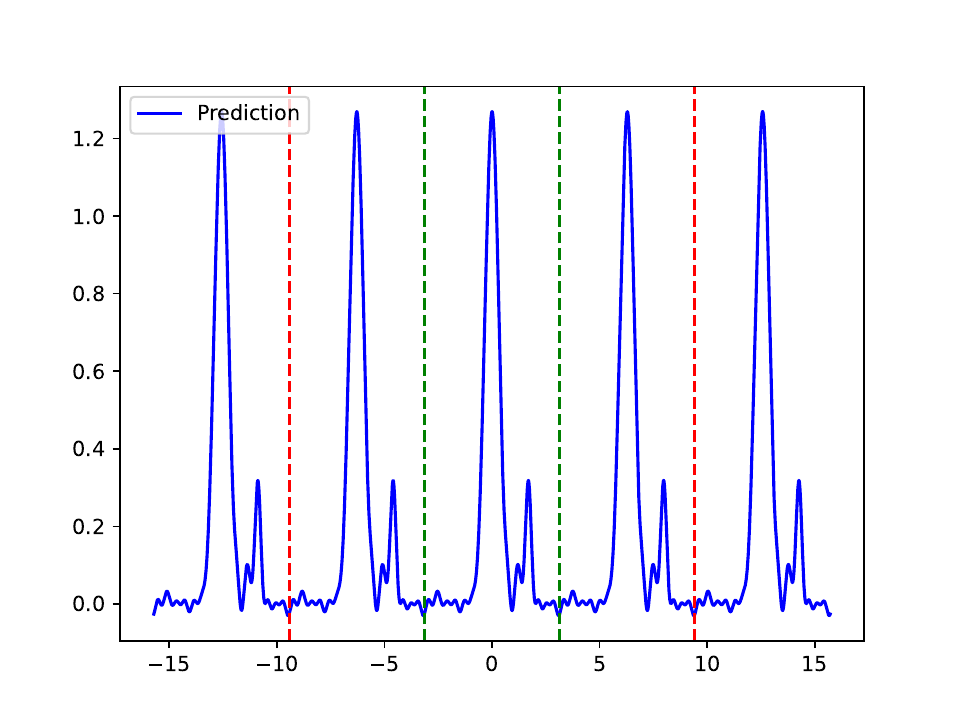}}
     \subfloat[Training in $\brackets{-2\pi, 2\pi}$
]{\includegraphics[width = 0.465\textwidth]{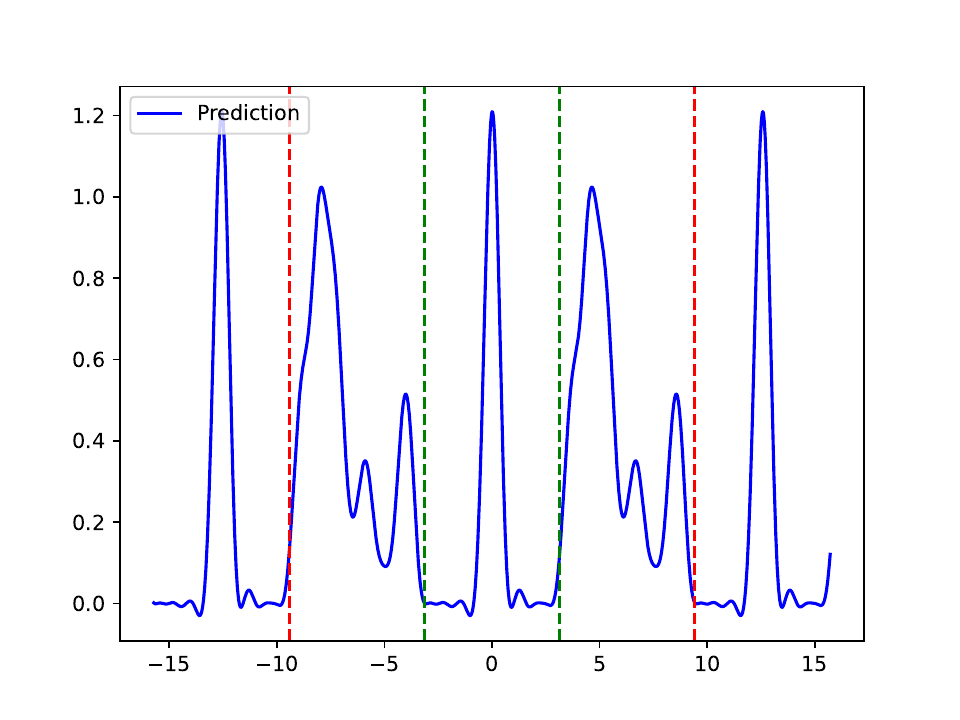}}
    \caption{Approximation of the PDF with training in different intervals.}
    \label{condist}
\end{figure}

\begin{figure}[H]
    \centering
     \subfloat[Training in $\brackets{-\pi, \pi}
$]{\includegraphics[width = 0.465\textwidth]{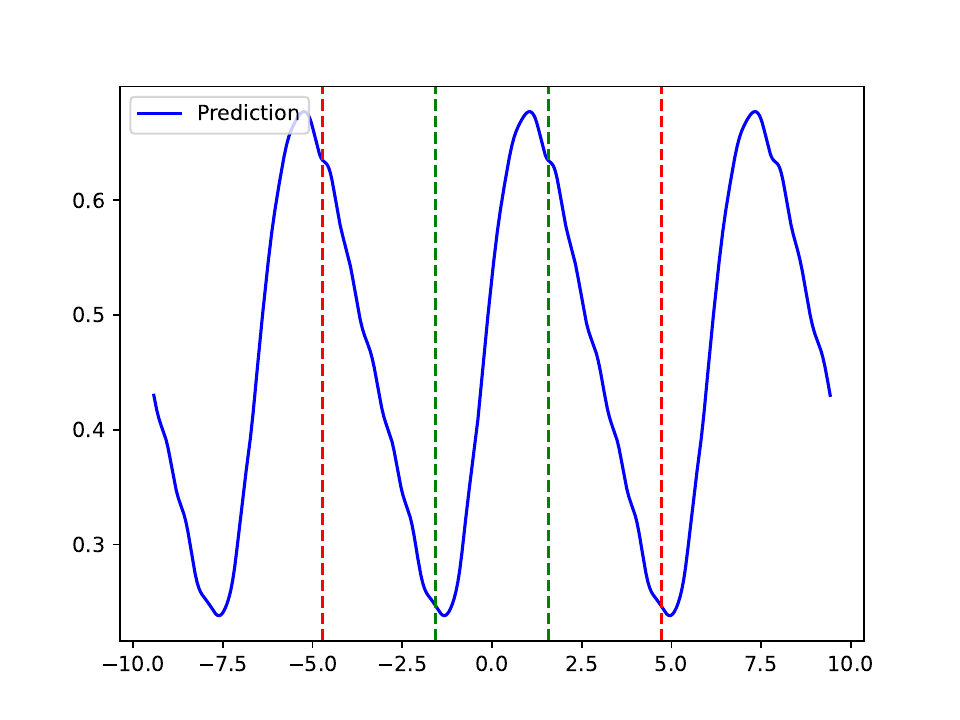}}
     \subfloat[Training in $\brackets{-2\pi, 2\pi}$
]{\includegraphics[width = 0.465\textwidth]{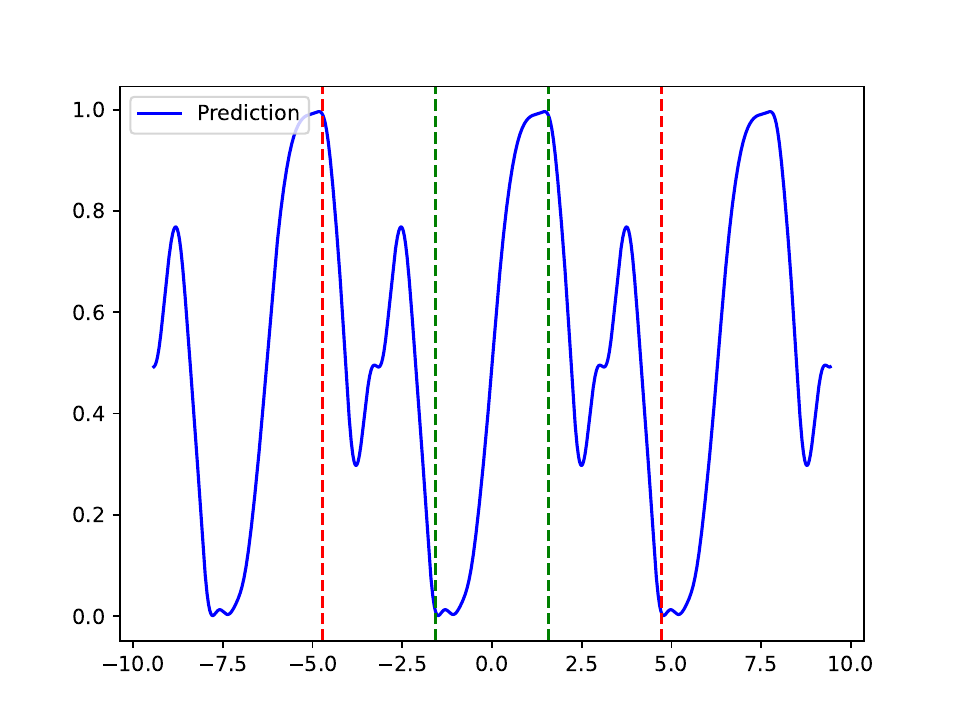}}
    \caption{Approximation of the CDF with training in different intervals.}
    \label{concdf}
\end{figure}

Another significant aspect that has to be taken into account is that the model to obtain the statistical functions is going to be trained in $[-2\pi, 2\pi]$ with data rescaled to $[-\pi, \pi]$, rather than trained directly in $[-\pi, \pi]$. This ensures that the resulting Fourier series is smoother and does not exhibit Gibbs phenomena, since outside $[-\pi, \pi]$ an approximation freedom is allowed due to the lack of data information outside that region. An example of the impact of this feature can be observed in Figure \ref{condist} but, specially, in Figure \ref{concdf}, because of the sharp jumps that the the CDF exhibits at its edges due to the periodic extension, causing the oscillations to become more pronounced and the approximation to be worse.

\begin{figure}[h]
\centering
\includegraphics[width=0.45\textwidth]{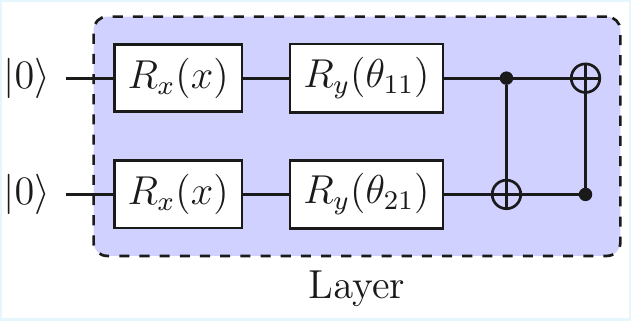}
\caption{Scheme of a single layer of the quantum \textit{ansatz} used to construct the PQC for 2 qubits, extracted from \cite{manzano2024thesis}.}
\label{esquemaansatz}
\end{figure}

In addition, all experiments related to these first two methods are conducted using the same quantum \textit{ansatz}, illustrated in Figure \ref{esquemaansatz}, which is composed of a fixed number of qubits and layers. However, this design can be scaled in complexity to analyze how the results vary with respect to the capacity of the circuit. This strategy allows to study both the accuracy and generalization of the model in different scenarios.

The last proposed method is based on the use of QAMC. As a QAE routine we select a modified version of the RQAE algorithm, termed mRQAE (see \cite{manzano_et_al_2025} for details), which allows to estimate not only the magnitude but also the sign of the quantity of interest. This is crucial in this case, since the Fourier coefficients can be either positive or negative. Thus, by employing the mRQAE algorithm, we will be able to obtain the Fourier coefficients of the underlying distribution to a certain precision and eventually calculate the derivative's price.

Although the structure of this third quantum method differs from the previously proposed ones, its primary purpose in this context is to serve as a benchmark for assessing the performance of the designed circuits. Specifically, it provides a set of reference numerical results that enable a meaningful comparison in terms of both cost and accuracy in the extraction of Fourier coefficients.

In order to establish this comparison, the QAMC method will be evaluated under various error tolerances achievable by Methods I and II. For each tolerance level, we will measure the number of circuits runs for each coefficient (interpreted as quantum samples in the context of QAMC). This number of executions will serve as a proxy for computational cost, thus providing an empirical cost metric against which the efficiency of the designed PQCs can be evaluated.

Next, we explain in more detail the relevant methodological aspects of the three mentioned approaches.

\subsubsection{Method I: Supervised Learning for PDF approximation}

First, we consider a labeled dataset associated with the probability distribution 
%\begin{equation*}
$\mathcal{T} = \{(x_i, y_i) \in Z = \mathcal{X} \times \mathcal{Y} \sim P(x, y) ~|~ \forall i \in \{0, \dots, I-1\} \}$,
%\end{equation*}
defined over a truncation interval that ensures the Fourier series approximation is sufficiently accurate. This interval is obtained following the same reasoning described in \cite{cosine}.

Next, we train the PQC using the empirical risk function defined in (\ref{ref_risk_generico}) applied to the PDF, i.e.,
\begin{align}
R_{h^1}^S(f^*, f) &=
\frac{1}{I} \sum_{i=0}^{I-1} \left( f^*(x_i) - f(x_i) \right)^2 + \frac{1}{I} \sum_{i=0}^{I-1} 
\left( \frac{d f^*}{dx}(x_i) - \frac{d f}{dx}(x_i) \right)^2,
\label{dml}
\end{align}
to approximate the underlying PDF in $[-2\pi, 2\pi]$ with data rescaled to $[-\pi, \pi]$, as mentioned before.

Subsequently, the Fourier series coefficients are extracted using a precise and efficient strategy based on the Discrete Fourier Transform (DFT), applied to direct evaluations of the trained circuit in $[-\pi, \pi]$.  
For this purpose, we use the scalar function that returns the output of the trained quantum circuit for each input $x$, corresponding to the Fourier series in exponential form, i.e.,
\begin{equation*}
    f(y|x) \approx \sum_{k=-\mathcal{K}}^{\mathcal{K}} c_k \cdot e^{i k x},
\end{equation*}
where $\mathcal{K}$ is the specified maximum degree and the coefficients $c_k \in \mathbb{C}$. To obtain the desired trigonometric form, the coefficients $A_k^f$ and $B_k^f$ of the trigonometric series are computed from the complex coefficients $c_k$ of the exponential form, using the identities
\begin{equation*}
    A_k^f = (c_k + c_{-k}), \quad B_k^f = -i (c_k - c_{-k}).
\end{equation*}
Then, the Fourier series coefficients associated with the payoff function, $C_k$ and $D_k$, are typically available in closed-form. Finally, the price is computed using the expression:
\begin{equation}
V(t_0,x)  \approx \frac{1}{2}(b - a) e^{-r\Delta t}\left( \frac{A_0^f \cdot C_0}{2} + \sum_{k=1}^\mathcal{K} \left( A_k^f \cdot C_k + B_k^f \cdot D_k \right) \right).
\label{precing}
\end{equation}

\subsubsection{Method II: Self-supervised Learning for CDF approximation}

As mentioned before, in derivative pricing, the exact probability distribution of the underlying asset is rarely available. Therefore, in this case, self-supervised learning will be employed, providing the model with only a representative set of asset samples, so that the labels required to construct the model’s cost function are generated internally from that set.

The formulation of the process is based on the results of \textit{Chapter} 7 of \cite{manzano2024thesis}, where, from various convergence results, an adaptation of the classical optimization problem is proposed. This starts from a target function $F^* \in \mathcal{T} \subseteq H^k$ mapping inputs $x \in \mathcal{X}$ to target labels $y \in \mathcal{Y}$, a model $F \in \mathcal{M} \subseteq H^k$, and a risk function $R_{H^k} : \mathcal{M} \longrightarrow \mathbb{R}^{+} \cup \{0\}$, where the goal is to find the best approximation $F$ of the target $F^*$ such that
\begin{equation*}
    F = \arg \min_{\hat{F} \in \mathcal{M}} R_{H^k}(\hat{F}).
\end{equation*}
The main differences introduced in this method are that the training dataset is given solely by samples and that the function to be estimated will be the CDF. For this purpose, we start by defining the dataset
%\begin{equation*}
    $\mathcal{T} = \{x_i \in \mathcal{X}~|~ x_i \sim F^*, \ i \in \{0, \ldots, I - 1\}\}$,
%\end{equation*}
where neither labels nor their derivatives are available. Therefore, it is necessary to define a new risk function that uses only the inputs $x \in \mathcal{X}$ and somehow incorporates the derivatives to ensure convergence of the approximation.

To do this, we first consider the case where the empirical risk $R$ is defined as the squared norm in the discretized space $ l^2(\mathcal{X}) $:
\begin{equation*}
R^{\mathcal{T}}_{l^2}(F) = \frac{1}{I} \sum_{i=0}^{I-1} (F^*(x_i) - F(x_i))^2.
\end{equation*}
Since in our case the true labels $ F^*(x_i) $ are not available, we approximate them by the empirical CDF:
\begin{equation*}
F^*(x) \approx F^*_{\text{emp}}(x) = \frac{1}{I} \sum_{i=0}^{I-1} \mathbf{1}_{x_i \leq x},
\end{equation*}
so that we use the empirical risk
\begin{equation*}
R^{\mathcal{X}}_{\mathcal{T}, l^2}(F) = \frac{1}{I} \sum_{i=0}^{I-1} \left( F^*_{\text{emp}}(x_i) - F(x_i) \right)^2.
\label{eq:riesgo_emp_cdf}
\end{equation*}
Next, we want to apply a similar procedure to the derivative of $F$, i.e., the probability density function $ f $. To this end, we consider the following risk:
\begin{equation*}
R_{L^2}(f) = \int_{\mathcal{X}} (f^*(x) - f(x))^2  \, dx = \int_{\mathcal{X}} f^*(x)^2 \, dx - 2 \int_{\mathcal{X}} f(x) f^*(x) \, dx + \int_{\mathcal{X}} f(x)^2 \, dx. \label{eq:riesgo_pdf}
\end{equation*}
Instead of working with the full expression of $R_{L^2}(f)$, we consider each term separately. Firstly, the term
\begin{equation*}
-2 \int_{\mathcal{X}} f(x) f^*(x) \, dx,
\end{equation*}
can be easily approximated by a Monte Carlo method:
\begin{equation*}
-2 \int_{\mathcal{X}} f(x) f^*(x) \, dx \approx -\frac{2}{n} \sum_{i=0}^{I-1} f(x_i).
\end{equation*}
Secondly, the term
\begin{equation*}
\int_{\mathcal{X}} f(x)^2 \, dx,
\end{equation*}
can be approximated by any numerical integration method $ Q $:
\begin{equation*}
\int_{\mathcal{X}} f(x)^2 \, dx \approx Q(f^2).
\end{equation*}
Finally, the term
\begin{equation*}
\int_{\mathcal{X}} f^*(x)^2 \, dx,
\end{equation*}
is a constant. If, in a minimization problem, we remove a constant from the function to be minimized, the minimum value of the function changes but not the point where this minimum is achieved, which in our case is the function $ f $. That is, we have
\begin{equation*}
f = \arg\min_{\hat{f} \in \mathcal{M}} R_{L^2}(\hat{f}) \Longleftrightarrow f = \arg\min_{\hat{f} \in \mathcal{M}} \left[ R_{L^2}(\hat{f}) - \|f^*\|^2_{L^2} \right].
\end{equation*}
Combining these ideas, the empirical risk we construct for the total risk based on the PDF takes the form
\begin{equation*}
R^{\mathcal{T}}_{l^2}(f) = -\frac{2}{I} \sum_{i=0}^{I-1} f(x_i) + Q(f^2). \label{eq:riesgo_emp_pdf}
\end{equation*}
In order to build a new empirical risk function in a similar way to (\ref{ref_risk_generico}), we can consider the combination of the empirical risks of $F$ and $f$:
\begin{equation*}
    R^{\mathcal{T}}_{l^2, l^2}(F) = R^{\mathcal{X}}_{\mathcal{T}, l^2}(F) + R^{\mathcal{T}}_{l^2}(f) = \frac{1}{I} \sum_{i=0}^{I-1} \left( F^*_{\text{emp}}(x_i) - F(x_i) \right)^2 - \frac{2}{I} \sum_{i=0}^{I-1} f(x_i) + Q(f^2),
\end{equation*}
which is related to the use of DML features for the CDF estimation.

Moreover, in order to guarantee that the CDF approximation is sufficiently accurate at the extremes of the function domain, we add an additional constraint term to the empirical risk function, resulting in
% \begin{equation*}
% 	R^{\mathcal{T}}_{l^2, l^2}(F) = \frac{1}{I} \sum_{i=1}^{I-2} \left( F^*_{\text{emp}}(x_i) - F(x_i) \right)^2 - \frac{2}{I} \sum_{i=0}^{I-1} f(x_i) + Q(f^2) + \left( F^*_{\text{emp}}(x_0)\right)^2 + \left( F^*_{\text{emp}}(x_{I-1}) - 1\right)^2.
% \end{equation*}
\begin{eqnarray*}
R^{\mathcal{T}}_{l^2, l^2}(F) &= &\frac{1}{I} \sum_{i=1}^{I-2} \left( F^*_{\text{emp}}(x_i) - F(x_i) \right)^2 \\
& &- \frac{2}{I} \sum_{i=0}^{I-1} f(x_i) + Q(f^2) \\
& &+ \left( F^*_{\text{emp}}(x_0)\right)^2 + \left( F^*_{\text{emp}}(x_{I-1}) - 1\right)^2.
\end{eqnarray*}
Since the sample sizes will be sufficiently representative, the constraints introduced here do not involve a forced approximation, but rather help to improve the fit at those points that may exhibit instabilities, particularly at the extremes. Thus, given an unlabeled dataset where the truncation interval is determined by the sample elements, the PQC will be trained in $[-2\pi,2\pi]$ but with data rescaled to $[-\pi,\pi]$. Subsequently, the series coefficients will be extracted using the DFT, applied to the direct evaluations of the trained circuit in $[-2\pi,2\pi]$. This allows us to recover the coefficients $c_k$ of the exponential series and turn them into in their trigonometric form.

Finally, as mentioned, the Fourier series coefficients associated with the payoff function are often obtained analytically, allowing to compute the final price using the expression 
% \begin{align*}
% V(t_0,x) \approx e^{-r\Delta t} \Bigg(  h(T,b)F(b) - h&(T,a)F(a) - \Big( \frac{A_0^F \cdot C_0^a}{2} + \sum_{k=1}^\mathcal{K} \left( A_k^F \cdot C_k^a + B_k^F \cdot D_k^a \right) \Big) - \\ &- \Big( \frac{A_0^F \cdot C_0^b}{2} + \sum_{k=1}^\mathcal{K} \left( A_k^F \cdot C_k^b + B_k^F \cdot D_k^b \right) \Big) \Bigg).
% \end{align*}
\begin{eqnarray*}
V(t_0,x) &\approx& e^{-r\Delta t} \left(  h(T,b)F(b) - h(T,a)F(a) \right) \\
& & - e^{-r\Delta t} \left( \frac{A_0^F \cdot C_0^a}{2} + \sum_{k=1}^\mathcal{K} \left( A_k^F \cdot C_k^a + B_k^F \cdot D_k^a \right) \right) \\
& & - e^{-r\Delta t} \left( \frac{A_0^F \cdot C_0^b}{2} + \sum_{k=1}^\mathcal{K} \left( A_k^F \cdot C_k^b + B_k^F \cdot D_k^b \right) \right)
\end{eqnarray*}
Note that the reason for using a Fourier series of period $2(b-a)$, as defined in (\ref{cdfseries}), in the interval
\begin{equation*}
    [\hat{a}, \hat{b}] = \left[\frac{3a-b}{2}, \frac{3b-a}{2}\right],
\end{equation*}
is that the the Gibbs phenomenon is completely eliminated, because we are only considering the part of the series corresponding to the CDF when defining the quantities in \eqref{eq:Ck_Dk} between $[a,c]$ and $[c,b]$, respectively. This is also why these are not the Fourier coefficients of the series that approximates $ h(T,y) $.

\subsubsection{Method III: QAMC with mRQAE for PDF approximation}

In this case, we propose an approach similar to the one employed in the first method, in the sense that we will compute the values of the coefficients of the trigonometric Fourier series of the underlying price PDF, although using QAMC.

We start from the same conditions as in Section  \ref{subsec31}, assuming the existence of an interval $[a, b] \subset \mathbb{R}$ and a number $\mathcal{K}$ of terms so that the Fourier series represents the PDF with sufficient accuracy, i.e.,
\begin{equation}
    f(y | x) \approx  \frac{A_0}{2} + \sum_{k=1}^\mathcal{K} \left( A_k^f \cos\left(2\pi k\frac{(y - a)}{b - a}\right) + B_k^f \sin\left(2\pi k\frac{(y - a)}{b - a}\right) \right),
    \label{seriprob}
\end{equation}

In order to compute the coefficients, we proceed as described in \cite{manzanomrqae}, restricting ourselves to models where an exact simulation of the asset's evolution can be performed, thereby avoiding errors arising from the use of numerical methods such as Euler-Maruyama. Therefore, when working for example with a Black-Scholes model, we can also assume the existence of a unitary operator that encodes the distribution of paths.

This results in the need for only a single register of $n\times n$ qubits to perform the entire simulation, so that we generate a set of $I$\footnote{Note that it is not necessarily true that $I=2^n$ for $n$ the number of qubits.} labeled data,
%\begin{equation*}
$\mathcal{T} = \{(S_i, f(S_i))\in  \mathcal{X} \times \mathcal{Y} ~|~  ~\forall i \in \{0, \dots, I-1\} \}$,
%\end{equation*}
that allows us to estimate the coefficients as
\begin{equation*}
A_k^f = \frac{2}{b-a} \int_{a}^{b} f(y | x) \cos\left(2\pi k\frac{(y - a)}{b - a}\right) \, dy \approx \frac{2}{b-a} \sum_{i=0}^{I-1} f(S_i)\, \cos\left(2\pi k\frac{(S_i - a)}{b - a}\right),
\end{equation*}
\begin{equation*}
B_k^f = \frac{2}{b-a} \int_{a}^{b} f(y | x) \sin\left(2\pi k\frac{(y - a)}{b - a}\right) \, dy \approx \frac{2}{b-a} \sum_{i=0}^{I-1} f(S_i)\, \sin\left(2\pi k\frac{(S_i - a)}{b - a}\right).
\end{equation*}

As mentioned, to achieve this we make use of the mRQAE, which is an asymptotically more efficient version of the RQAE. Their main difference is that parameters such as the confidence and the required precision in each iteration are chosen following different criteria. Note again that this QAE routine enables to recover the sign of the coefficients and compute the price accurately. 

Again, we assume that the Fourier series coefficients associated with the payoff function, $C_k$ and $D_k$, can be evaluated via analytical expressions, enabling to calculate the derivative's price using (\ref{precing}).

\section{Numerical results}\label{sec4}

In the following section, we present and discuss the results obtained from the different experiments, highlighting the main trends and insights derived from the comparative evaluation of the proposed methods.

\subsection{Experiment setting}\label{subsec41}

For our numerical experiments, we consider the pricing of a European vanilla put option at $t_0=0$, so that the payoff takes the form (see \cite{hull}, for example):
\begin{equation*}
    h(T,S_T) = \max\{K - S_T, 0\}.
\end{equation*}
In order to price an option, the stochastic dynamics of the underlying asset price $S_t$ has to be introduced, usually in terms of a stochastic differential equation (SDE). In the present work, we assume that the underlying asset follows a Black–Scholes-type dynamics, which is described by the following SDE:
\begin{equation}
  dS_t = r S_t \, dt + \sigma S_t \, dB_t, 
\label{bs_SDE}
\end{equation}
where $S_0$ is the given price of the underlying at $t=0$, $r$ is the risk-free interest rate, $\sigma$ is the constant volatility of the assets, and $B_t$ represents a standard Brownian motion under the probability measure $\mathbb{Q}$.

In the case of more complex dynamics, such as stochastic volatility models, the simulation of the underlying asset prices to obtain $S_T$ (and therefore $h(T,S_T)$) requires the use of numerical methods to solve SDEs. However, in the case of Black-Scholes dynamics, the exact solution of SDE (\ref{bs_SDE}) can be obtained from Ito calculus and is given by
\begin{equation}
S_t=S_0 \exp\left(\left(r-\frac{\sigma^2}{2}\right)t+B_t \right).
\label{GMB_t0}
\end{equation}
It should be remarked that, in Black–Scholes models for asset prices, when valuing derivatives it is common to work with logarithmic normalized by the strike price,  due to the transformation of the statistical behavior of the underlying asset. Note that expression (\ref{GMB_t0}) implies that the model assumes the following lognormal distribution for the asset price at time $t$: 
\begin{equation*}
    S_t\sim \text{Lognormal}(r, \sigma^2t),
\end{equation*}
then it can be easily proven that the logarithmic normalized prices $X_t$ follow a normal distribution as follows: 
\begin{equation}
    X_t := \log(S_t / K) \sim \mathcal{N} \left( \log\left( \frac{S_0}{K} \right) + \left( r - \frac{1}{2} \sigma^2 \right) t,\ \sigma^2 t \right).
    \label{log_prices}
\end{equation}
The goal of this transformation is to work with a symmetric and unbounded distribution, such as the normal distribution, which is particularly useful in contexts where techniques based on Fourier theory, machine learning, or quantum simulation are employed, since many of these tools operate more naturally and efficiently in symmetric domains centered around zero.

Note that when working with the process $X_t$, the consideration of its probability distribution (\ref{log_prices})  allows one to obtain the expression of the payoff $h$ and the PDF $f$ that appear in the integral expression (\ref{option_price}).

The following model parameters have been chosen:
\begin{equation*}
	S_0 = 100,\quad r = 0.1,\quad T = 1,\quad \sigma = 0.25.
\end{equation*}
These values allow the simulation of a realistic yet controlled market scenario, suitable for evaluating the generalization ability and accuracy of the proposed methods. Additionally, three different strike prices have been selected:
\begin{equation*}
	K = 90, ~~ K = 100 ~~ \text{and} ~~ K = 110,
\end{equation*}
with the aim of analyzing the model's behavior for different option contract configurations where the spot price $S_0$ is above, equal to, or below the strike price $K$, respectively.

Regarding the PQC training setup, the hyperparameters used in the different experiments are summarized in Table \ref{tab:params_comparacion}.

\begin{table}[H]
    \caption{Training hyperparameters used in methods I and II.}
	\label{tab:params_comparacion}
	\centering
	\begin{tabular}{|c|c|c|}
		\hline
		\textbf{Hyperparameter}    & \textbf{I} & \textbf{II} \\ \hline
		Optimizer              & Adam        & Adam       \\ \hline
		Learning rate              & $0.005$        & $0.1$       \\ \hline
		Epochs                     & $300$          & $300$       \\ \hline
		Supervised weight          & $0.9$          & $0.2$       \\ \hline
		Differential weight        & $0.1$          & $0.8$       \\ \hline
		Training points            & $250 - 2.5\cdot 10^3$          & $10^3 - 10^4$     \\ \hline
		Test points                & $100$          & $10^3$      \\ \hline
		Repetitions               & $10$           & $10$        \\ \hline
	\end{tabular}
\end{table}

Additionally, experiments have been conducted with different regular $(n\times n)$ configurations (referring to the $n$ qubits and $n$ layers employed in the PQC), to assess whether another number of parameters improves the results. However, as it is common in QML problems, there is neither universal optimal configuration, nor a clear relationship between scalability and accuracy. In many cases, it is necessary to find an appropriate balance (a trade-off) between model the complexity and the specific characteristics of the problem under consideration. For the (QAMC-based) Method III, the number of coefficients used will be the same as in Method I, by construction, and we vary the number of executions of the circuit, often called \emph{shots} in the quantum jargon, by prescribing an increasing tolerance for the mRQAE routine.

In general, the structures used to design the PQCs in the experiments are chosen such that they return accurate approximations of the distributions and the payoff and capture the complexity of the problem. For Method I, the minimum scheme is $6\times6$, while for Method II it is $4\times4$.

Finally, \textit{Table} \ref{tab:entorno} shows the technical characteristics of the computational system for the experiments.

\begin{table}[H]
    \caption{Technical characteristics of the computational environment used.}
	\label{tab:entorno}
	\centering
	\begin{tabular}{|c|c|}
		\hline
		\textbf{Parameter} & \textbf{Value} \\ \hline
		Processor & Intel(R) Core(TM) i7-8550U CPU @ 1.80GHz \\ \hline
		RAM Memory & 8.0 GB \\ \hline
		Operating System & Windows 10 (64 bits) \\ \hline
		Python Version & 3.12.7 
		\\ \hline
		JAX Version & 0.4.35 \\ \hline
		PennyLane Version & 0.40.0 \\ \hline
	\end{tabular}
\end{table}

\subsection{Results and discussion} 

\begin{figure}[H]
    \centering
    \subfloat[$6\times6$]{%
        \includegraphics[width=0.465\textwidth]{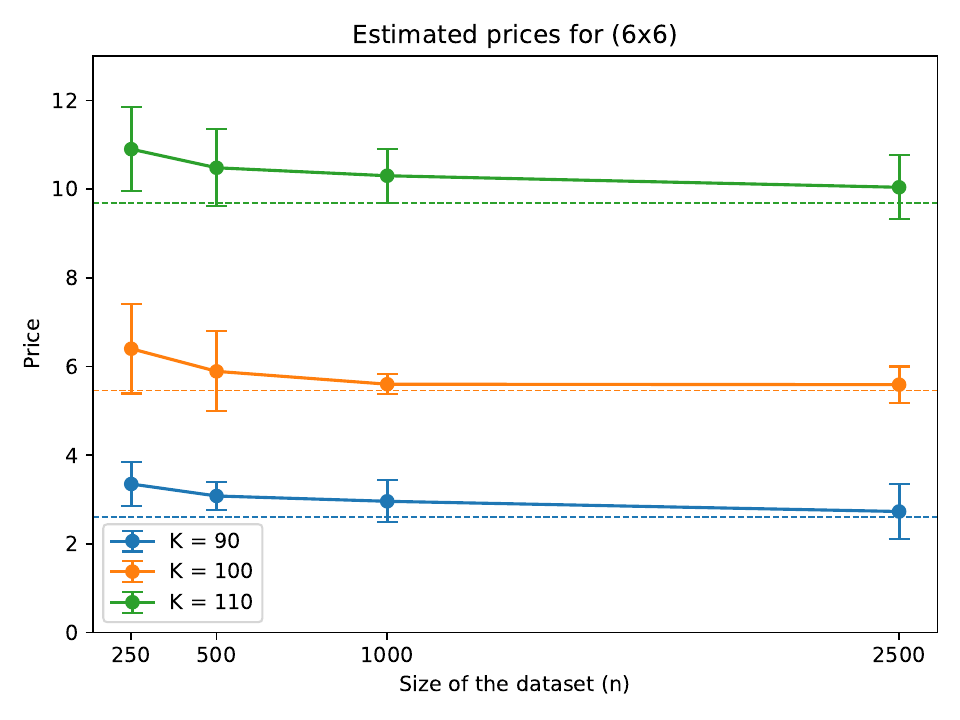}%
    }
    \centering
    \subfloat[$4\times4$]{%
        \includegraphics[width=0.465\textwidth]{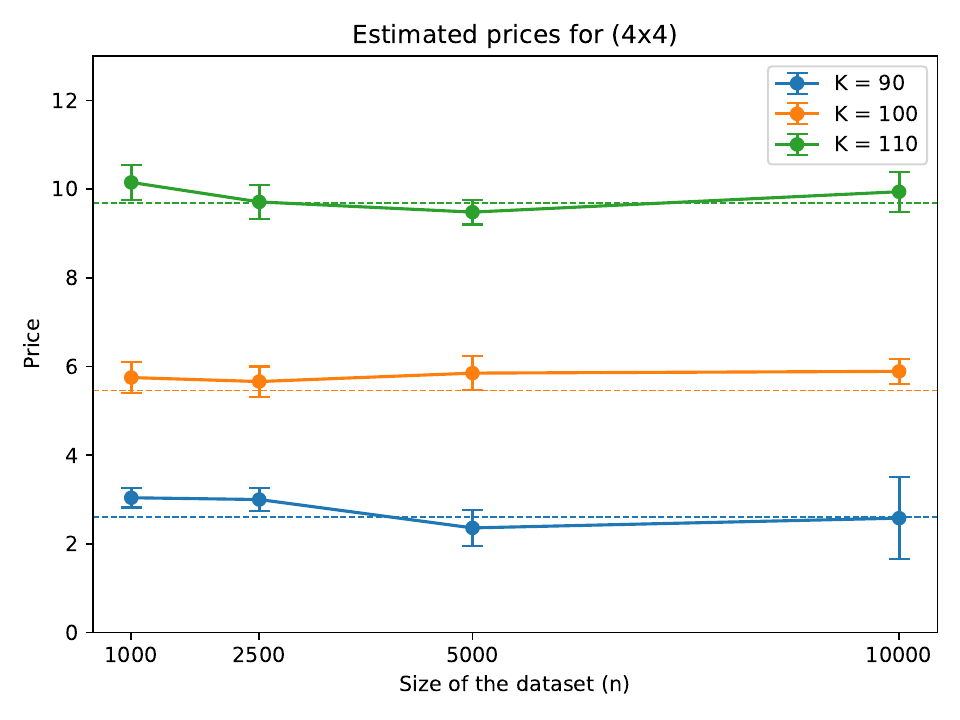}%
    }
    \\
    \subfloat[$7\times7$]{%
        \includegraphics[width=0.465\textwidth]{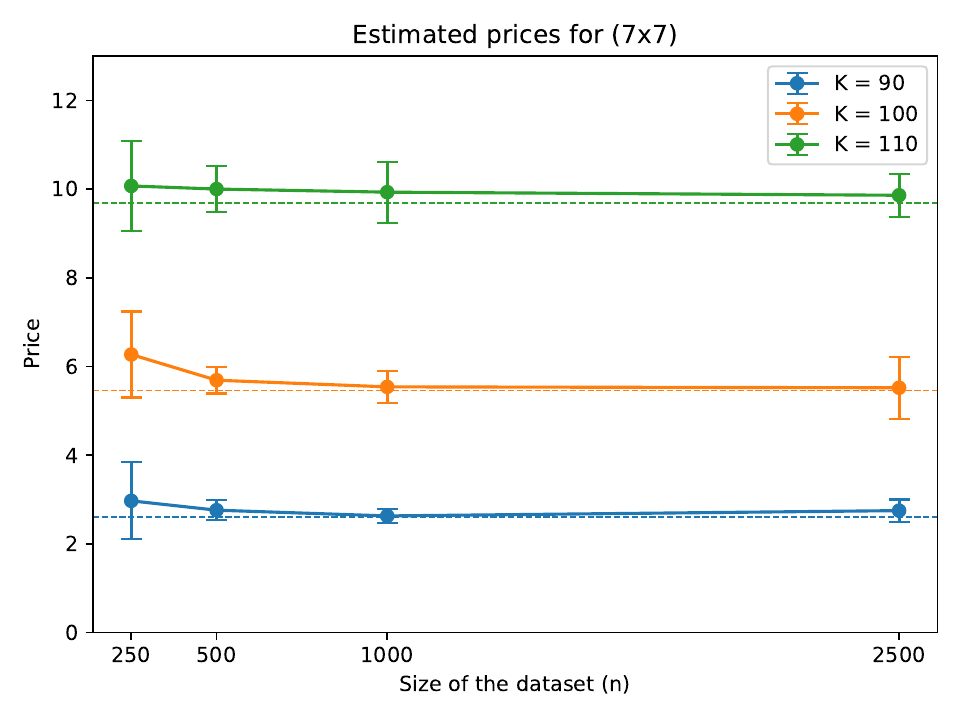}%
    }
    \subfloat[$5\times5$]{%
        \includegraphics[width=0.465\textwidth]{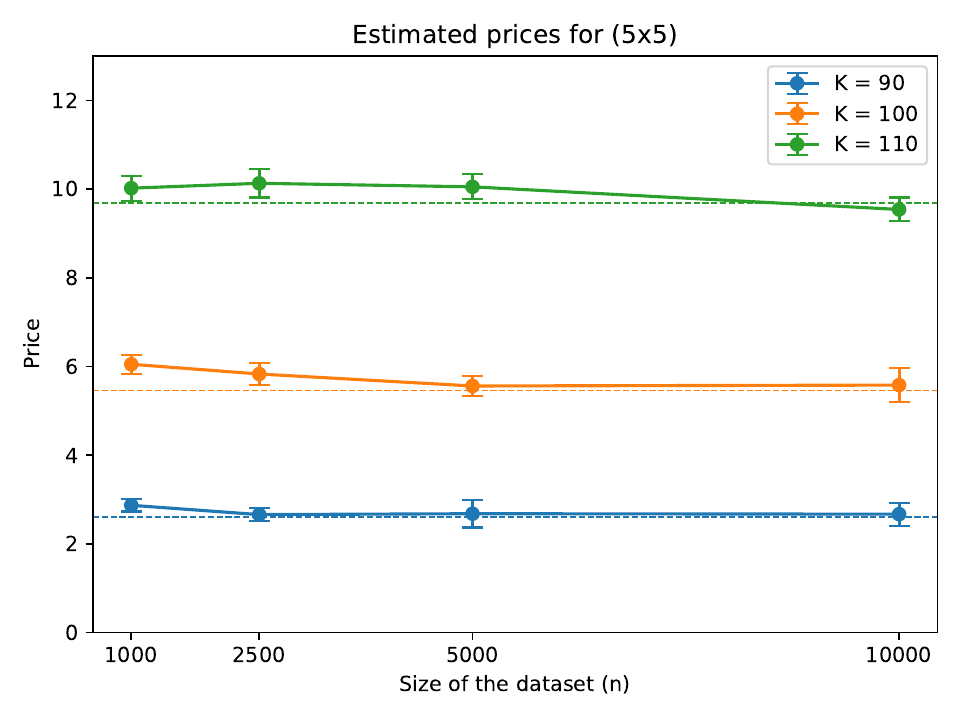}%
    }
    \\
    \subfloat[$8\times8$]{%
        \includegraphics[width=0.465\textwidth]{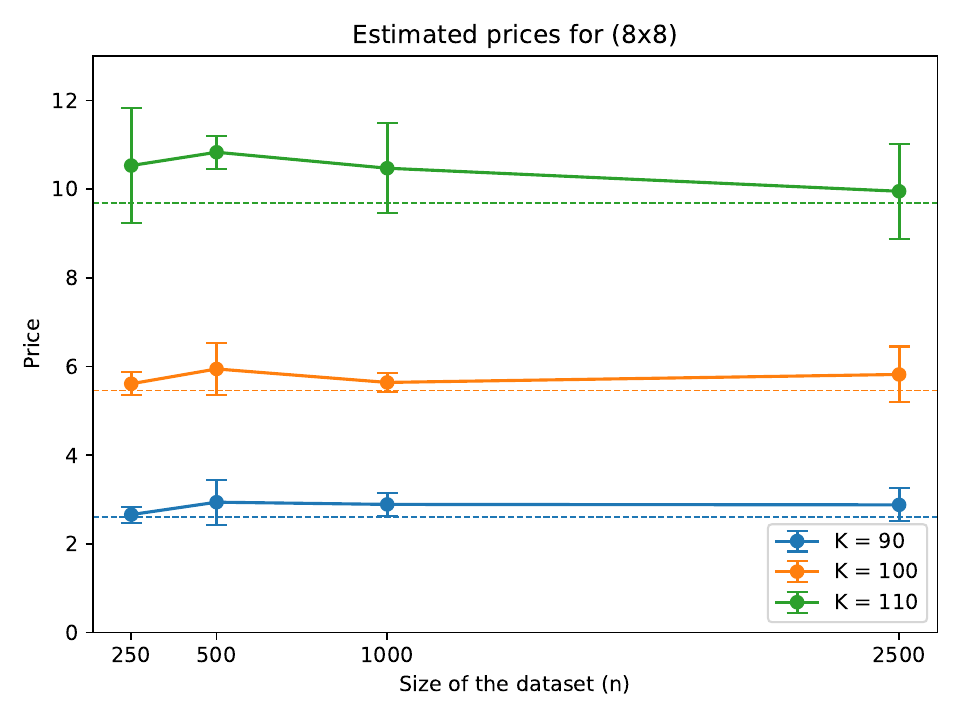}%
    }
    \subfloat[$6\times6$]{%
        \includegraphics[width=0.465\textwidth]{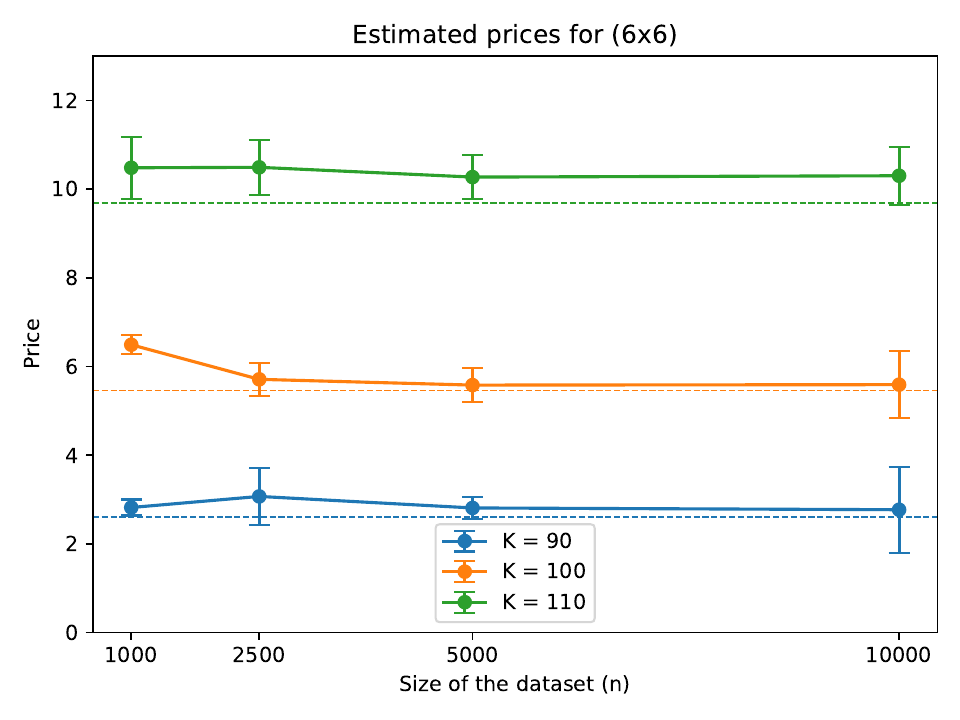}%
    }
    \caption{Convergence results for Method I (left column) and Method II (right column).}
    \label{fig:convergence_MI_MII}
\end{figure}

In this section, we firstly present the convergence results obtained by the three methods described in Section \ref{sec3}. Several experiments have been carried out for each method with an increasing approximation power of the PQC and the results are shown in Figure \ref{fig:convergence_MI_MII} (Method I and Method II) and Figure \ref{fig:convergence_MIII} (Method III), respectively. Each of the pictures in these graphs show the accuracy convergence in terms of the size of each case's employed dataset for the three considered strike prices. Next, in Sections \ref{subsec451} and \ref{subsec452}, we discuss the presented results more in depth, analyzing the impact on the PQC estimations in terms of the dataset size and the PQC structures (number of coefficients in the approximation).

% \begin{figure}[H]
%     \centering
%     \subfloat[$6\times6$]{%
%         \includegraphics[width=0.45\textwidth]{precios_PDF_Lig_6x6.pdf}%
%     }
%     \subfloat[$7\times7$]{%
%         \includegraphics[width=0.45\textwidth]{precios_PDF_Lig_7x7.pdf}%
%     }
%     \\
%     \subfloat[$8\times8$]{%
%         \includegraphics[width=0.45\textwidth]{precios_PDF_Lig_8x8.pdf}%
%     }
%     \caption{Convergence results for Method I.}
%     \label{fig:convergence_MI}
% \end{figure}

% \begin{figure}[H]
%     \centering
%     \subfloat[$4\times4$]{%
%         \includegraphics[width=0.45\textwidth]{precios_CDF_Lig_4x4.pdf}%
%     }
%     \subfloat[$5\times5$]{%
%         \includegraphics[width=0.45\textwidth]{precios_CDF_Lig_5x5.pdf}%
%     }
%     \\
%     \subfloat[$6\times6$]{%
%         \includegraphics[width=0.45\textwidth]{precios_CDF_Lig_6x6.pdf}%
%     }
%     \caption{Convergence results for Method II.}
%     \label{fig:convergence_MII}
% \end{figure}

% \begin{figure}[H]
%     \centering
%     \subfloat[$6\times6$]{%
%         \includegraphics[width=0.465\textwidth]{price_III_6x6_v2.pdf}%
%     }
%     \subfloat[$7\times7$]{%
%         \includegraphics[width=0.465\textwidth]{price_III_7x7_v2.pdf}%
%     }
%     \\
%     \subfloat[$8\times8$]{%
%         \includegraphics[width=0.465\textwidth]{price_III_8x8_v2.pdf}%
%     }
%     \caption{Convergence results for Method III.}
%     \label{fig:convergence_MIII}
% \end{figure}

\subsubsection{Impact of the data size}\label{subsec451}

One of the main objectives of increasing the sample size is to obtain a more complete and representative dataset, in order to capture the underlying distribution in a more faithful way. As the number of observations increases, the estimations obtained through approximation models should reflect more accurately the statistical patterns of the original data, thus resulting in a better generalization and a reduction of the error.

From Figure \ref{fig:convergence_MI_MII}, a clear convergence towards the exact value in both Method I and Method II can be observed as the sample size increases. However, some outliers appear, which can be attributed to the inherent error of the training process, particularly to the randomness in the initialization of the quantum model weights, which introduces fluctuations that are not always corrected during optimization. These aspects highlight the importance of carrying out a larger number of experimental repetitions, accompanied by a statistical analysis that isolates these sources of uncertainty.

\begin{figure}[H]
    \centering
    \subfloat[$6 \times 6$]{\includegraphics[width=0.33\textwidth]{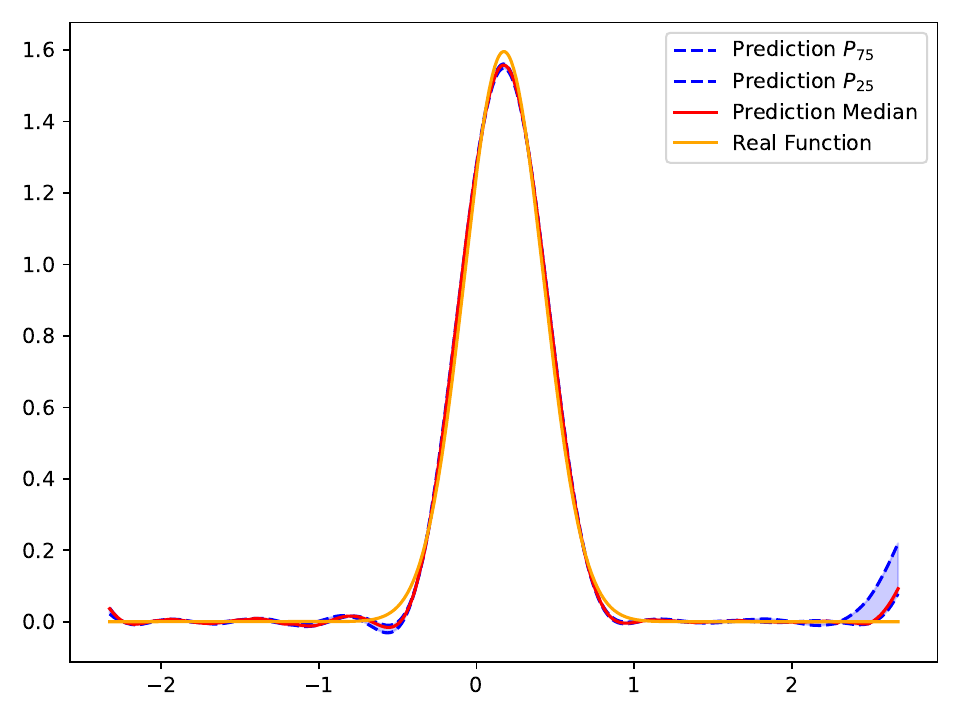}}
    \subfloat[$7 \times 7$]{\includegraphics[width=0.33\textwidth]{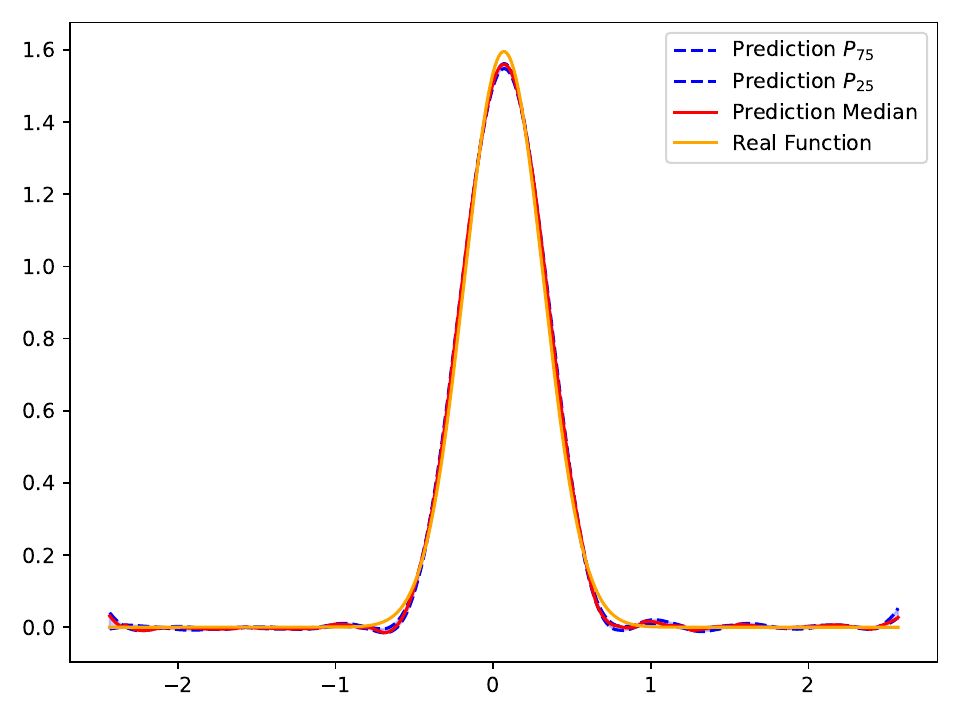}}
    \subfloat[$8 \times 8$]{\includegraphics[width=0.33\textwidth]{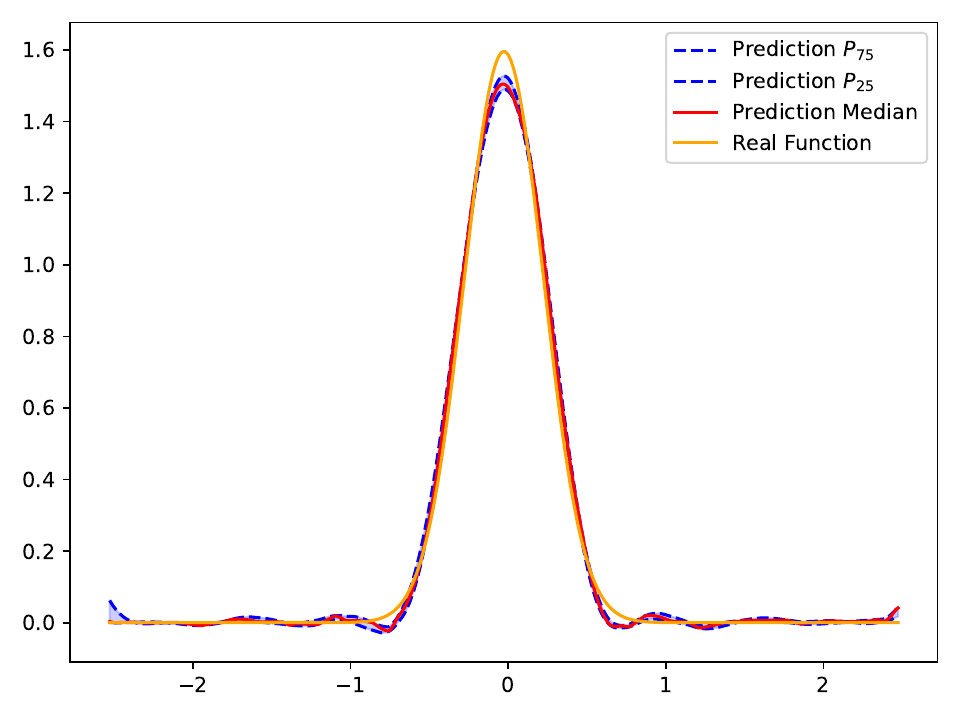}}
    \caption{Output of the PQC for Method I (a) and for Method II (b).}
    \label{resultito1}
\end{figure}

\begin{figure}[H]
    \centering
    
    \subfloat[$4\times 4$]{\includegraphics[width=0.33\textwidth]{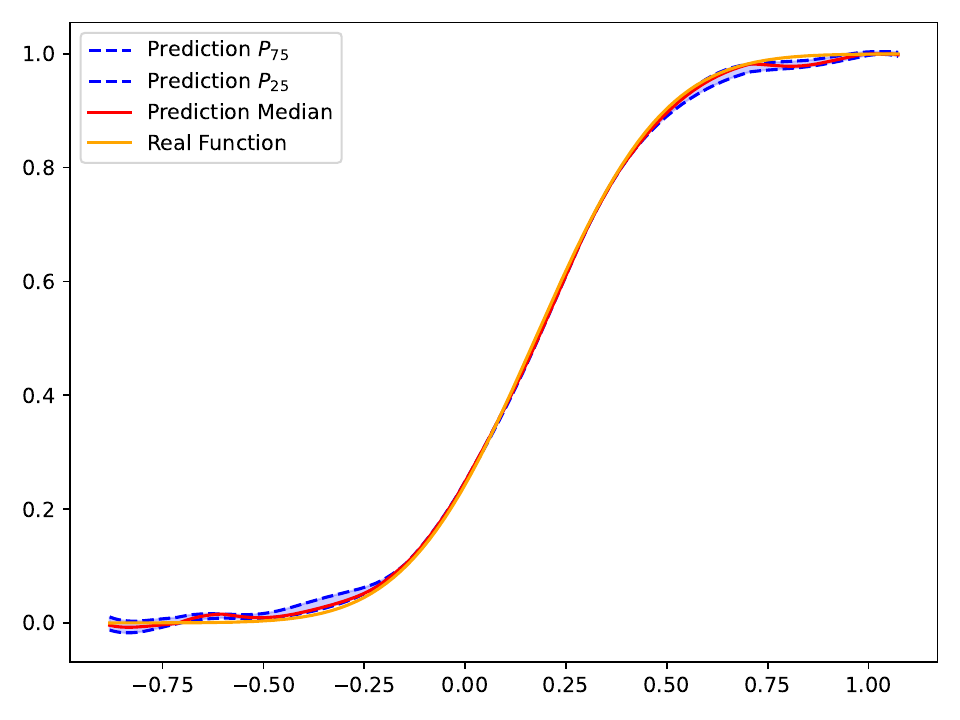}}
    \subfloat[$5 \times 5$]{\includegraphics[width=0.33\textwidth]{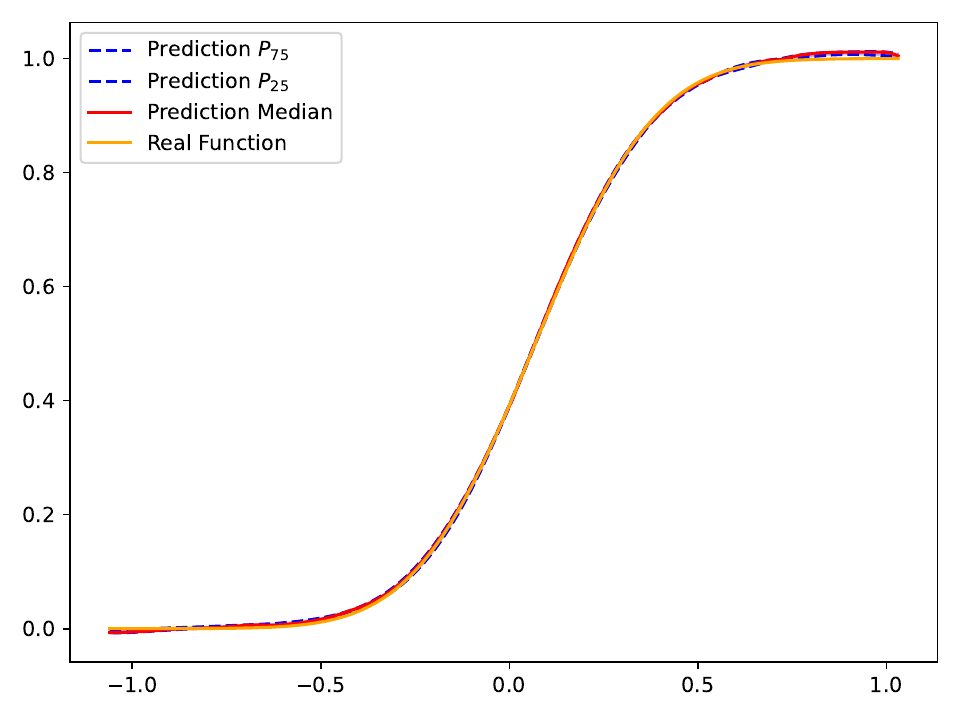}}
    \subfloat[$6 \times 6$]{\includegraphics[width=0.33\textwidth]{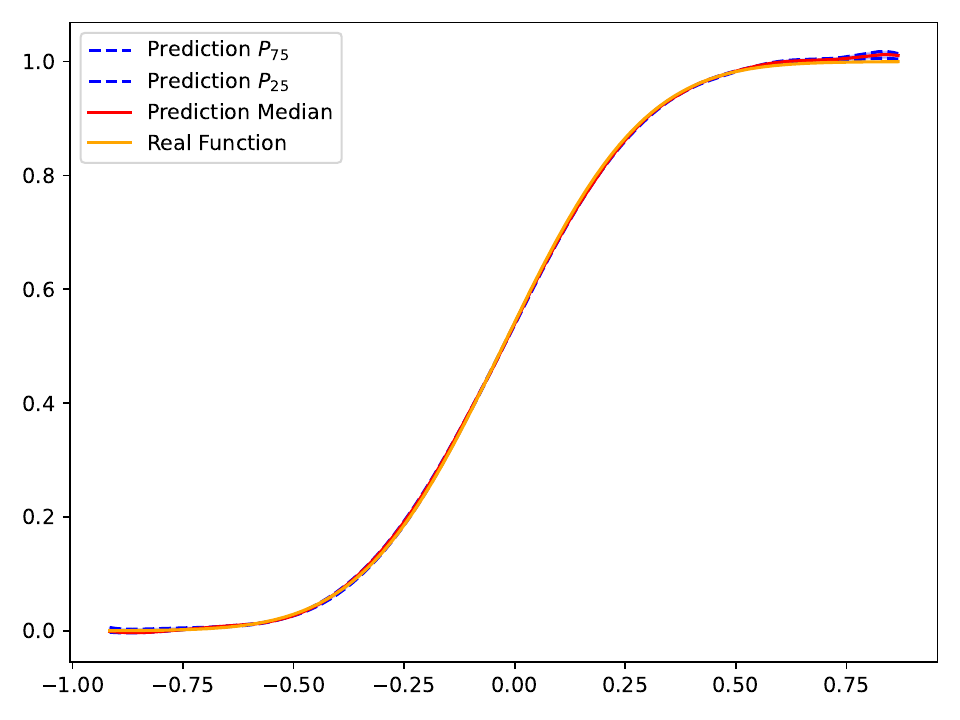}}
    \caption{Output of the PQC for Method II.}
    \label{resultito2}
\end{figure}

Despite these limitations, the results obtained so far are clearly satisfactory and demonstrate the potential of the approach, as illustrated in Figures \ref{resultito1} and \ref{resultito2} for Method I and Method II, respectively.

% \begin{figure}[H]
%     \centering
%     \subfloat[$6 \times 6$]{\includegraphics[width=0.365\textwidth]{pdf_90_resultado.pdf}}
%     \subfloat[$4\times 4$]{\includegraphics[width=0.365\textwidth]{cdf_90_resultado.pdf}}
%     \caption{Output of the PQC for Method I (a) and for Method II (b).}
%     \label{resultito1}
% \end{figure}

% \begin{figure}[H]
%     \centering
%     \subfloat[$7 \times 7$]{\includegraphics[width=0.365\textwidth]{pdf_100_resultado.pdf}}
%     \subfloat[$5 \times 5$]{\includegraphics[width=0.365\textwidth]{cdf_100_resultado.pdf}}
%     \caption{Output of the PQC for Method I (a) and for Method II (b).}
%     \label{resultito2}
% \end{figure}

% \begin{figure}[H]
%     \centering
%     \subfloat[$8 \times 8$]{\includegraphics[width=0.365\textwidth]{pdf_110_resultado.pdf}}
%     \subfloat[$6 \times 6$]{\includegraphics[width=0.365\textwidth]{cdf_110_resultado.pdf}}
%     \caption{Output of the PQC for Method I (a) and for Method II (b).}
%     \label{resultito3}
% \end{figure}

\begin{figure}[H]
    \centering
    \subfloat[$6\times6$]{%
        \includegraphics[width=0.465\textwidth]{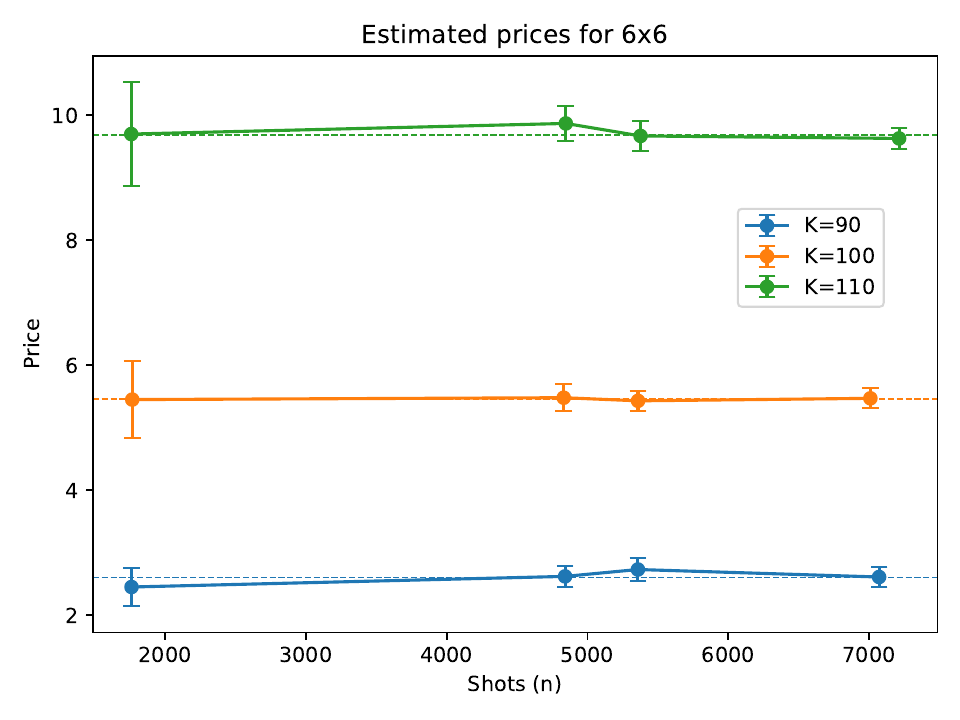}%
    }
    \\
    \subfloat[$7\times7$]{%
        \includegraphics[width=0.465\textwidth]{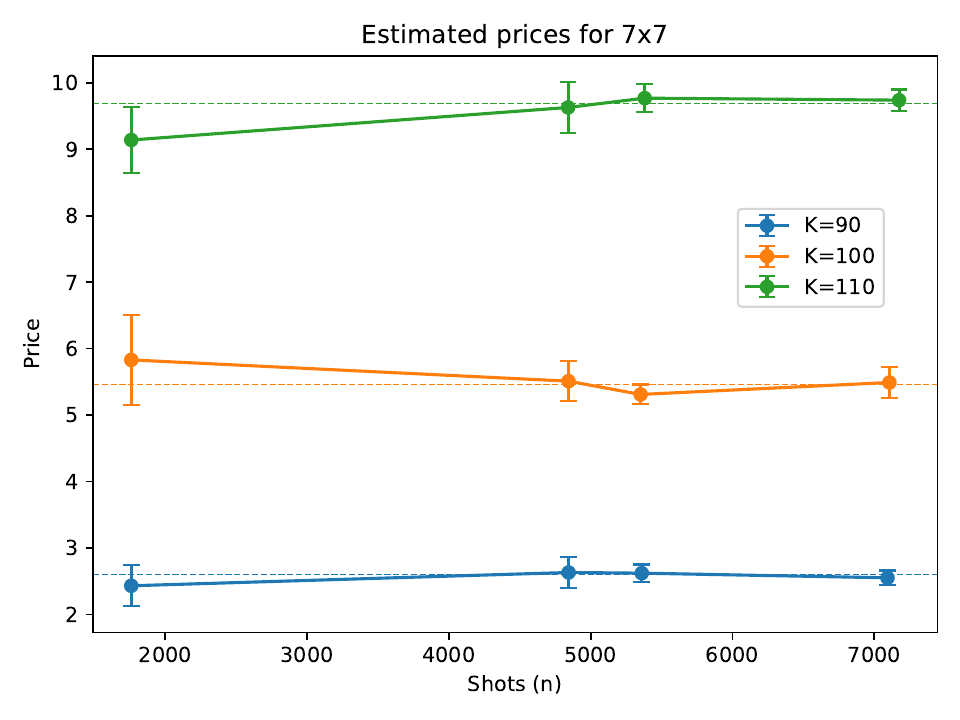}%
    }
    \\
    \subfloat[$8\times8$]{%
        \includegraphics[width=0.465\textwidth]{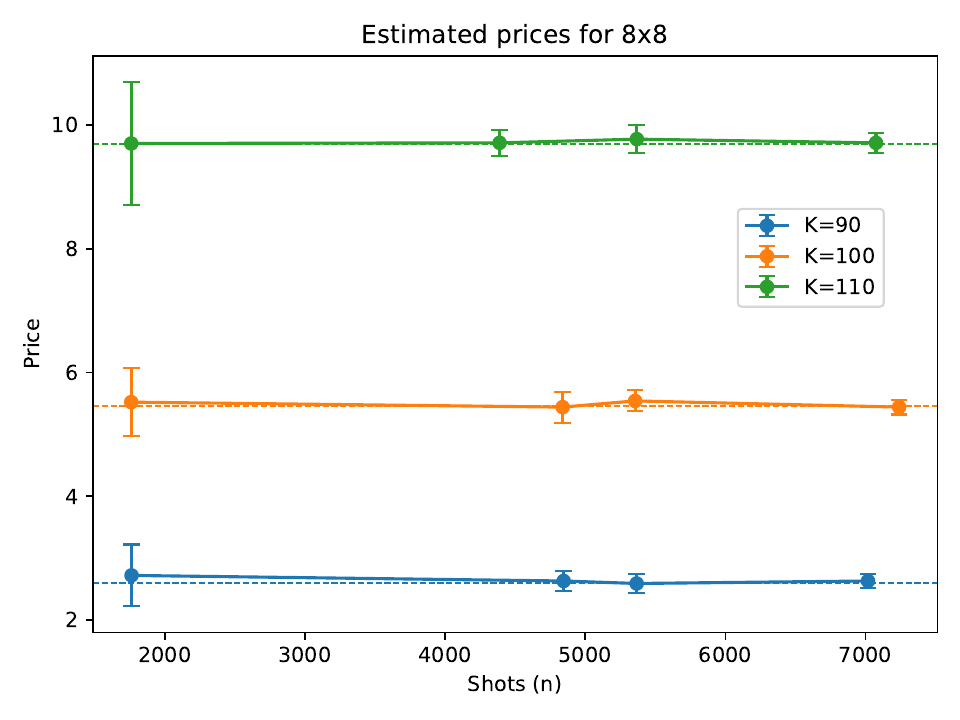}%
    }
    \caption{Convergence results for Method III.}
    \label{fig:convergence_MIII}
\end{figure}

The Method III (Figure \ref{fig:convergence_MIII}) achieves remarkably accurate estimations, with reduced variability (specially when more shots are employed). Those effects are somehow expected since, while the precision and variability of the estimations provided by the PQCs (Methods I and II) are mainly affected by the quality of the optimization process (subject to the intrinsic variability of initial parameters, random jumps, etc.) and the data, the precision by QAMC is controlled by the QAE routine (mRQAE), prescribing that the discrete data points are enough to accurately represent the continuous density function. 

Note however that, in order to get highly accurate predictions, we need around $5000$ shots per coefficient in average, thus implying an order of hundreds of thousand shots to obtain the final price. In comparison, with Method II we obtain similar results with $10000$ data samples (and, of course, after a training procedure), which highlights the potential of QML-based approaches in the derivatives valuation area, as a complementary methodology to the more popular QAMC. From a practical point of view, the performance of Method II is even more relevant, since it does not require the availability of the underlying PDF, but only the generation of random samples following the proper distribution.

\subsubsection{Impact of the number of coefficients}\label{subsec452}

Since the functions to be approximated in each method are different, it is necessary to use distinct dimensions in the PQC schemes to ensure good results. This is because, in certain cases, adding an excessive number of terms in the approximation of a function can introduce additional contributions to oscillations and instabilities in the model. In the specific case of the CDF, which exhibits sharp jumps at its boundaries due to periodic extension, these oscillations may become more pronounced.  

Moreover, the exponential nature of the payoff and the periodicity of the approximation give rise to abrupt jumps at the points of discontinuity. As a result, the outcomes may present larger errors and significant variability as the number of terms in the Fourier expansion increases. 

Nevertheless, when using dimensions on the order of $6\times6$, $7\times7$, and $8\times8$ for Method I, and $4\times4$, $5\times5$, and $6\times6$ for Method II, both methods display stable behavior, showing a clear trend of convergence towards the true derivative value as the expressive capacity of the circuit increases. This can be clearly seen in Figure \ref{fig:convergence_MI_MII}, reinforcing the idea that introducing flexibility in the Fourier series approximation is essential to achieve accurate and reliable estimates.  

It is worth noting that, according to the results, the best performance is obtained with an intermediate scheme ($7\times7$ for Method I and $5\times5$ for Method II), illustrating how in this type of problem there is often a \textit{trade-off} between model complexity and the specific characteristics of the problem under study.  
On the other hand, in the case of Method III, increasing the number of coefficients leads to a substantial improvement in the results, particularly when the number of shots is low. Beyond a certain threshold, the coefficients tend to take significantly small values and, combined with the accuracy provided by QAMC, it clearly shows how these small contributions help to refine the computed price, thus illustrating a clear convergence of the estimated prices.

% \begin{figure}[H]
%     \centering
%     \subfloat[$6\times6$]{%
%         \includegraphics[width=0.465\textwidth]{price_III_6x6_v2.pdf}%
%     }
%     \\
%     \subfloat[$7\times7$]{%
%         \includegraphics[width=0.465\textwidth]{price_III_7x7_v2.pdf}%
%     }
%     \\
%     \subfloat[$8\times8$]{%
%         \includegraphics[width=0.465\textwidth]{price_III_8x8_v2.pdf}%
%     }
%     \caption{Convergence results for Method III.}
%     \label{fig:convergence_MIII}
% \end{figure}

\section{Conclusions}\label{sec5}

In this work, we have presented two hybrid classical--quantum approaches for the option pricing problem. This framework connects quantum learning models with Fourier-based valuation techniques, enabling the extraction of relevant statistical information from quantum-generated data. The performance of the proposed models was benchmarked against QAMC, allowing for a quantitative comparison in terms of computational cost and accuracy in the recovery of Fourier coefficients. The number of circuit executions per coefficient, interpreted as quantum samples, served as an empirical cost metric to evaluate PQC efficiency.

Both methods exhibited stable convergence towards the true derivative value as the expressive capacity of the PQCs increased. Circuit dimensions on the order of $6\times6$ to $8\times8$ for Method I and $4\times4$ to $6\times6$ for Method II yielded consistent and reliable estimates, confirming the importance of model flexibility in Fourier-based approximations.

In addition, as the sample size is increased, a systematic convergence towards the exact solution has been observed across methods, though minor outliers appeared due to stochastic effects in the training phase, particularly from random initialization of quantum weights. These effects suggest the need for multiple experimental runs and statistical post-analysis to isolate uncertainty sources.

Finally, Method II obtains comparable precision using only $\sim 10000$ data samples as the benchmarking Method III, which requires on the order of hundreds of thousands of shots to achieve high accuracy with low variability. This fact underscores the potential of QML-based strategies as efficient and complementary alternatives to traditional QAMC schemes. Moreover, Method II stands out for not requiring explicit knowledge of the underlying probability density function, instead relying on sampling from the appropriate distribution, as it is common practice in industry.

% \appendix
% \section{An example appendix} 
% \lipsum[71]

\section*{Acknowledgements}
All authors acknowledge the support of CITIC, as a center accredited for excellence within the Galician University System and a
member of the CIGUS Network, receiving subsidies from the Department of Education, Science, Universities, and Vocational Training of the Xunta de Galicia. Additionally, it is co-financed by the EU through the FEDER Galicia 2021-27 operational program (ref. ED451G 2023/01). Á. Leitao and C. Vázquez acknowledge the funding from the Ministry of Science and Innovation of Spain (ref. PID2022-141058OB-I00) and from the Department of Education, Science, Universities, and Vocational Training of the Xunta de Galicia (ref. ED451C 2022/047), both including FEDER financial support. Á. Leitao acknowledges the financial support from the Spanish Ministry of Science and Innovation through the Ramón y Cajal 2022 grant, and the Department of Education, Science, Universities, and Vocational Training of the Xunta de Galicia through the Excellence research program (ref. ED431F 2025/032).

\printbibliography

\end{document}